\documentclass[a4paper,fleqn]{cas-dc} 
\ExplSyntaxOn
\cs_set:Npn \__first_footerline: {\mbox{}}
\ExplSyntaxOff

\ExplSyntaxOn
\cs_gset:Npn \__first_footerline:
  { \group_begin: \small \sffamily \__short_authors: \group_end: }
\ExplSyntaxOff 

\usepackage[numbers,sort&compress]{natbib}
\usepackage{threeparttable}

\usepackage{pdflscape}
\usepackage{longtable}
\usepackage{booktabs}
\usepackage{array}
\usepackage{multirow}
\usepackage{graphicx}
\usepackage{wrapfig}
\usepackage{makecell}
\usepackage{tabularx}
\usepackage{multicol}
\usepackage{calc}
\usepackage{enumitem}
\usepackage{xurl}
\usepackage{tabularx}
\usepackage{hyperref}
\usepackage{tikz}
\usetikzlibrary{shapes.geometric, arrows.meta, positioning}
\usepackage{array}
\usepackage{float}
\usepackage{caption}
\usepackage{cuted}
\usepackage{placeins}
\usepackage{microtype}
\newcolumntype{C}[1]{>{\centering\arraybackslash}m{#1}}
\newcolumntype{L}[1]{>{\raggedright\arraybackslash}m{#1}}

\definecolor{mylightblue}{HTML}{00CCFF}
\definecolor{mydeepblue}{HTML}{0A03DB}

\begin{document}
\let\WriteBookmarks\relax
\def\floatpagepagefraction{1}
\def\textpagefraction{.001}
\shorttitle{Ancillary-Service Markets Review}
\shortauthors{A.C. Makrides et~al.}

\title [mode = title]{Ancillary Services in High-Renewable Power Systems: Comparing Market Design, Emerging Trends, and Future Challenges}




\author[1]{A.C. Makrides}[type=editor,
                        auid=000,bioid=1,
                        orcid=0009-0001-1795-0729]
\cormark[1]
\ead{andreas.makrides24@imperial.ac.uk}

\credit{Conceptualisation, Data curation, Investigation, Formal analysis, Validation, Visualisation, Writing – original draft}

\affiliation[1]{organization={Dyson School of Design Engineering, Imperial College London},
                addressline={Exhibition Rd, South Kensington}, 
                city={London},
                postcode={SW7 2AZ}, 
                country={United Kingdom}}

\author[1]{A. Churkin}[orcid=0000-0002-7875-9178]
\ead{a.churkin@imperial.ac.uk}
\credit{Supervision, Writing – review and editing}

\author[2]{F. Teng}[orcid=0000-0002-6828-0294]
\ead{f.teng@imperial.ac.uk}
\credit{Supervision, Writing – review and editing}

\affiliation[2]{organization={Department of Electrical and Electronic Engineering, Imperial College London},
                addressline={Exhibition Rd, South Kensington}, 
                city={London},
                postcode={SW7 2AZ}, 
                country={United Kingdom}}

\author[1]{P. Pinson}[orcid=0000-0002-1480-0282]
\ead{p.pinson@imperial.ac.uk}
\credit{Supervision, Writing – review and editing}


\cortext[cor1]{Corresponding author}


\begin{abstract}
Rapid growth in renewable generation is changing power-system operation and increasing the importance of ancillary services for security and reliability. As inverter-based resources displace conventional synchronous generation, system operators must procure a broader set of services spanning five categories: frequency control and reserves, voltage and reactive power support, system stability, restoration, and congestion management. This paper compares ancillary-service market design in Great Britain, Germany, Denmark, Italy, California, Texas, and Australia. It examines how these systems define, procure, remunerate, and coordinate services as variable renewable generation increases. The review develops a consistent product-level classification, provides comparable timing visualisations of frequency-control and reserve products, and maps implementation modalities across services and systems. Four trends emerge: faster and more granular frequency-response products, more explicit procurement of stability capabilities including inertia, system strength, and dynamic voltage support, broader participation by storage, demand-side resources, and other inverter-based resources, and greater use of locational, targeted, and technology-neutral procurement. Challenges include limited harmonisation, incomplete valuation of flexibility and stability capabilities whose value depends on location and operating conditions, barriers to participation by distributed and emerging resources, and insufficient coordination between market and operational arrangements. The comparison finds no convergence towards a single market design and supports differentiated but coordinated ancillary-service frameworks reflecting each service's technical, temporal, and locational characteristics.
\end{abstract}



\begin{keywords}
Ancillary services \sep Electricity market design \sep Flexibility \sep Frequency control \sep Inverter-based resources \sep Power system stability \sep Renewable energy integration 
\end{keywords}

\maketitle

\section{Introduction}
\label{s1}

Variable and uncertain renewable energy sources (RES) are expanding rapidly across electricity systems worldwide, reshaping both system planning and real-time operation \cite{InternationalEnergyAgencyIEA2023COP28Gap, fes2025}. At the same time, energy storage and demand-side flexibility are becoming more important for balancing variability across multiple timescales \cite{McPherson2018,Schmidt2023MonetizingStorage}. However, the displacement of conventional synchronous generation by inverter-based resources (IBRs), such as wind, solar PV, and battery energy storage systems (BESS),  is also changing the physical behaviour of power systems \cite{Pfenninger2015RenewablesSecurity,10444680}. This shift has significant operational consequences. Lower system inertia leads to faster frequency deviations, while declining short-circuit levels reduce system strength and can make voltage control, fault detection, and post-disturbance recovery more challenging \cite{10508461,8779818,MAHMUD2016582,8757997}. These changes are increasing the need for explicitly defined and procured system-support capabilities \cite{pierluigistabilitty,8450880,9796617}.

Ancillary services have therefore become central to secure and resilient power-system operation \cite{roleofastoencourage}. Historically, synchronous generators supplied many of these services inherently or while operating part-loaded or in standby. Under high renewable penetration, this can become inefficient because conventional units may need to remain online primarily to provide system support rather than energy \cite{ba79c137-4f4a-3dbc-a69e-702be297425f}. Emerging technologies, including grid-forming IBRs and synchronous condensers, can provide both traditional services and newer capabilities such as fast frequency response and synthetic inertia \cite{JOHNSON2020114492,ESIG2025,Lasseter2020}. Yet market arrangements have not always evolved at the same pace as these technological capabilities \cite{9855885}. Procurement mechanisms, remuneration rules, and grid-code requirements often still reflect systems dominated by conventional generation, creating barriers to participation and incomplete incentives for essential system-support services \cite{Rancilio2022}.

These challenges differ across jurisdictions because ancillary-service design reflects system-specific regulatory frameworks, operational needs, market structures, reform priorities, system-operator practices, and historical development pathways \cite{Rancilio2022}. Existing studies often focus on individual services, single regions, or selected aspects of market design \cite{LOBATOMIGUELEZ2008515,MING201483,PRAKASH2022112303, BORNE2018605,8864014,robinepri}. A direct cross-system comparison remains limited, particularly under increasing renewable penetration and IBR deployment. Differences in terminology, product definitions, procurement mechanisms, remuneration rules, and participation requirements further complicate the identification of common trends and transferable lessons.

This paper addresses this gap through a document-based comparative review of ancillary-service market design across seven electricity systems: Great Britain (GB), Germany (DE), Denmark (DK), Italy (IT), Texas (TX), California (CA), and Australia (AU). The comparison is organised at the level at which market rules and system-operation arrangements are defined. Great Britain, Germany, Denmark, and Italy are therefore considered through their national or system-level frameworks, while Texas, California, and Australia are represented through ERCOT, CAISO, and the National Electricity Market, respectively. Where appropriate, the systems are referred to through their main system or market operators: the National Energy System Operator (NESO), the four German transmission system operators (50Hertz, TenneT Germany, Amprion, and TransnetBW), Energinet, Terna, ERCOT, CAISO, and the Australian Energy Market Operator (AEMO).

The review adopts a qualitative approach, systematically comparing official documents and relevant academic literature across five dimensions: (i) service definition and classification, (ii) procurement and remuneration, (iii) market structure and coordination, (iv) technology participation and access, and (v) recent reforms and market evolution.

The comparison reveals four persistent challenges:
\begin{itemize}[leftmargin=3em, label=\raisebox{0.2ex}{\large$\bullet$},
                itemsep=1pt, topsep=10pt, parsep=0pt, partopsep=0pt] 
    \item \textbf{Challenge \#1:} Limited harmonisation, particularly for non-frequency services, where product definitions and procurement approaches remain strongly system-specific,
    \item \textbf{Challenge \#2:} Incomplete valuation of flexibility and stability-related capabilities, especially when their contribution depends on location and operating conditions,
    \item \textbf{Challenge \#3:} Barriers to the participation of distributed and emerging resources, arising from technical requirements, qualification procedures, and market-access arrangements, and
    \item \textbf{Challenge \#4:} Weak coordination between energy, balancing, ancillary services, and network arrangements, which can limit the efficient allocation of flexibility across different system needs.
\end{itemize}

It also identifies four main market-design trends:
\begin{itemize}[leftmargin=3em, label=\raisebox{0.2ex}{\large$\bullet$},
                itemsep=1pt, topsep=10pt, parsep=0pt, partopsep=0pt]
    \item \textbf{Trend \#1:} Faster and more granular frequency-control and reserve products, reflecting the growing need for rapid response in systems with declining synchronous inertia,
    \item \textbf{Trend \#2:} More explicit procurement of stability-related capabilities, including inertia, system strength, and other services that were traditionally provided implicitly by synchronous generation,
    \item \textbf{Trend \#3:} Broader participation by emerging and non-traditional resources, supported by increasingly capability- and performance-based access requirements, and
    \item \textbf{Trend \#4:} Increasing use of locational and targeted procurement where service value depends strongly on network conditions and the location of the providing resource.
\end{itemize}

These findings show that ancillary-service markets are not converging towards a single design. Instead, procurement and remuneration arrangements continue to reflect the physical characteristics of each service, the underlying market architecture, and local system needs. The paper makes three main contributions. First, it develops a consistent product-level classification framework for ancillary-service products, mapping heterogeneous service definitions across the seven electricity systems into a five-category structure and comparable service characteristics. Second, it provides consistent timing visualisations that enable direct comparison of the response time and sustainment duration of frequency-control and reserve products across the reviewed systems. Third, it maps the implementation modalities used across service categories and electricity systems, distinguishing between mandatory or operational provision, targeted or contractual arrangements, and explicit or integrated market mechanisms. Together, these contributions provide a structured basis for identifying the cross-market trends, persistent challenges, and design implications discussed in the paper.

The remainder of the paper is organised as follows. Section~\ref{s2} introduces the role and classification of ancillary services in renewable-dominated power systems. Section~\ref{s3} presents the comparative approach adopted in this study. Section~\ref{s4} compares ancillary-service market design across the selected systems and operator frameworks. Section~\ref{s5} discusses the main challenges, trends, and market-design implications emerging from the comparison. Finally, Section~\ref{s6} concludes the paper.

\section{Ancillary Services in Renewable-Dominated Power Systems}
\label{s2}

This section introduces the operational role of ancillary services, their classification into five categories, and their evolving requirements under increasing renewable and IBR penetration.

\subsection{Role of Ancillary Services in Power System Operation}

Ancillary services refer to a set of operational functions required to maintain secure and reliable power-system operation beyond the physical delivery of electrical energy to end users \cite{roleofastoencourage, kirschen2026fundamentals}. They regulate system frequency, voltage, current, and power flows to keep these parameters within acceptable limits under both normal and post-disturbance conditions \cite{4077135, ancillartservicess}.

Furthermore, from an operational perspective, ancillary services provide system operators (SOs) the tools needed to balance supply and demand in real time, manage network constraints and congestion, and maintain an adequate response to contingencies \cite{ancillartservicess, powercontrol}. Their importance increases as the power system becomes more dynamic, less predictable, and more dependent on IBRs. Unlike energy markets, which primarily remunerate the scheduled delivery of electrical energy, ancillary services are tied directly to the physical and dynamic behaviour of the system. They may need to respond within seconds, sustain delivery over longer intervals, or provide continuous control depending on the operational need \cite{4077135}. Their value, therefore, depends not only on available capacity but also on speed, accuracy, location, duration, and compliance with defined performance requirements \cite{kibry2007, 6698583}.

As power systems transition towards higher shares of renewable and inverter-based generation, ancillary services are increasingly being specified, procured, and remunerated explicitly. This shift reflects their distinct operational role and the need to secure system-support capabilities across a wider and more uncertain range of operating conditions.

\subsection{Classification of Ancillary Services}
\label{s22}

Ancillary services are not classified in the same way across electricity systems. Terminology, product definitions, procurement arrangements, and technical requirements vary, reflecting differences in system needs, market design, and regulatory practice. For the purposes of this review, we group ancillary services into five broad system-level categories: frequency control and reserve services, voltage and reactive power support services, system stability services, restoration, and congestion management services \cite{4077135, entsoe,voltagecontrol,11415609, 5211165, LAMADRID20121959}.

\subsubsection{Frequency Control and Reserve Services}

Frequency control and reserve services maintain system frequency close to its nominal value, typically 50 or 60 Hz, by supporting the continuous balance between supply and demand \cite{4077135}. They respond to both normal system imbalances and larger disturbances, such as generator outages or sudden changes in demand or renewable output, and are essential for arresting, stabilising, and restoring frequency after such events \cite{entsoe, Eto2011FrequencyResponse}. Frequency control is usually organised as a sequence of layered actions that differ in response time, duration, and operational purpose. Fast frequency support acts almost immediately to arrest the initial frequency deviation. Primary control then contains and stabilises frequency within seconds, while secondary control restores frequency towards its nominal value and releases primary reserves. Tertiary control, also referred to as replacement reserve in some systems, restores the reserve margin over longer timeframes and prepares the system for subsequent contingencies \cite{SCHERER2013292}. Frequency and reserve services provide the flexible active-power capability needed to deliver these responses. They can be supplied by increasing generation, reducing demand, discharging storage, or otherwise adjusting active-power injections depending on system conditions and resource capabilities \cite{5608535, 7038111}.

\subsubsection{Voltage and Reactive Power Support Services}

Voltage and reactive power support services maintain voltage levels within planning and operational limits across the network \cite{4077135,voltagecontrol, powercontrol}. Providers of these services must be able to inject or absorb reactive power in response to local network conditions. Unlike frequency, which is uniform across an interconnected synchronous area, voltage is inherently locational and varies across buses, feeders, and network regions. Reactive power therefore has limited transferability and must be provided close to where voltage support is needed. This locational nature makes voltage support strongly dependent on network topology, power-flow patterns, and the availability of suitable resources at specific nodes in the system \cite{7271077, 917280}.

\subsubsection{System Stability Services}

System stability services support the ability of the power system to withstand, control, and recover from disturbances while maintaining stable operation under transient and dynamic conditions. Their importance is increasing as IBRs replace synchronous generation, because IBRs do not inherently provide the same stabilising characteristics as synchronous machines \cite{8450880, JOHNSON2020114492, DREIDY2017144}. As a result, capabilities that were historically embedded in conventional generation are increasingly being treated as distinct services that may need to be specified, procured, and remunerated explicitly.

Inertia is one of the most important stability-related capabilities because it limits the rate of change of frequency (RoCoF) immediately after a disturbance \cite{8779818}. By slowing the initial frequency decline, inertia gives frequency-control actions more time to respond before frequency reaches its nadir, i.e., the lowest frequency point following the event \cite{entsoeinertia}. System strength is another critical requirement and is commonly assessed using measures related to short-circuit level (SCL) \cite{nrecscl}. It affects voltage control, fault ride-through performance, and the correct operation of protection systems \cite{nerc2017}. As renewable penetration increases and synchronous generation is displaced, SCL levels may decline in affected network areas, reducing the system’s ability to manage faults and increasing the risk of protection maloperation or poor post-fault recovery \cite{pierluigistabilitty}. Dynamic voltage support (DVS) and other fast-acting stability services address these challenges by supporting voltage recovery and damping post-fault oscillations in voltage, power, and frequency \cite{962413, 8556012}. More broadly, system stability services capture an expanding set of capabilities required to maintain secure operation under low-inertia and weak-grid conditions, reflecting the evolving technical needs of modern power systems.

\subsubsection{Restoration Services}

Restoration services support system recovery following a partial or complete blackout \cite{11415609, 8362717}. They enable the re-energisation of the power system by allowing generation units and other capable resources to restart without relying on an external electricity supply \cite{blackstartrev}. These services support the staged restoration plan of the network and help SOs restore supply, reconnect demand, and return the system to a secure operating state after major disturbances \cite{LIANG2024109494, jainfunct}. Restoration services are therefore essential for system resilience, but they are typically event-driven and procured in advance from resources with specific technical capabilities and suitable locations.

\subsubsection{Congestion Management Services}

Congestion management services support secure system operation under network-constrained conditions \cite{11415609}. They address local or regional transmission constraints by keeping power flows within operational limits. These services can reduce the volume and cost of corrective actions required to manage network constraints, limit the impact of unforeseen events, and help defer or avoid network reinforcement where operational solutions are sufficient \cite{11415609, 5772585}. Unlike frequency-control services, congestion management is strongly location-dependent because the value of a resource depends on its position in the network and its ability to relieve specific constraints.

\subsection{Evolving Ancillary-Service Requirements in High-Renewable Systems}

Ancillary-service requirements are becoming more demanding, time-critical, and location-specific as power systems operate with higher shares of variable renewable generation and inverter-based resources \cite{ancillartservicess, RATNAM2020109773, 10.1093/ce/zkae105}. System operators increasingly require faster and more accurate responses to manage frequency and voltage deviations under more dynamic operating conditions. At the same time, capabilities that were previously supplied implicitly as a by-product of conventional generation are being defined and procured more explicitly to ensure their availability when synchronous generation is not online \cite{11415609}. This has led to the development of new or more differentiated services, including fast frequency response and inertia-related products, designed to address emerging stability and flexibility needs \cite{RAPIZZA2020100407,FERNANDEZMUNOZ2020109662}. In parallel, the set of potential service providers is expanding. BESS, demand-side resources, aggregated distributed energy resources (DERs), and other emerging providers are increasingly able to participate in ancillary-service markets and provide flexibility and system-support capabilities \cite{8864014,11415609}. 

These developments are driving a shift towards more dynamic, granular, and technology-neutral ancillary-service frameworks that better reflect the operational needs of modern power systems \cite{designancillary}. However, the appropriate balance between market-based procurement and mandatory technical requirements remains an open design question \cite{Rancilio2022}. Market-based products can reveal value, support competition, and encourage innovation but may struggle to capture highly locational or system-specific needs. Grid-code requirements can ensure minimum technical capability but may also create implicit technology preferences or reduce the visibility of service value. In practice, many systems are likely to require a combination of both approaches. The key challenge is therefore to define services with enough granularity to reflect actual system needs while avoiding excessive product complexity, over-specification, or regulatory barriers to technology-neutral participation \cite{ancillartservicess, designancillary}. This study does not seek to resolve this design trade-off. Instead, it compares how different electricity systems define, procure, and adapt ancillary services as operational requirements evolve.

\section{Comparative Approach}
\label{s3}

This study adopts a document-based framework to compare ancillary-service market design across the seven electricity systems introduced in Section~\ref{s1}. Comparative analysis is particularly useful in this context because ancillary services are shaped by both technical drivers (e.g., renewable integration and declining synchronous generation) and system-specific factors (e.g., regulatory frameworks and network characteristics). These systems were selected for their advanced electricity markets, increasing renewable penetration, active ancillary-service reform, and diverse institutional and operational arrangements \cite{ieaforecast2026, ENTSOE2025MarketReview}.

Great Britain is included as a dynamic and innovative market, shaped in part by its relatively isolated system, increasing renewable penetration, and limited synchronous interconnection with neighbouring systems. These characteristics have increased the importance of low-inertia, voltage, and system-strength requirements and have supported the early introduction of targeted frequency, voltage, and stability services \cite{neso_as}. Germany and Denmark represent systems operating within the broader European market framework, combining participation in harmonised European balancing platforms with national efforts to develop and integrate more specialised ancillary services \cite{germany_as, denmark_as}. Italy is included as a system closely aligned with established European structures, where ancillary-service design is evolving progressively in response to operational needs and regulatory developments \cite{terna_as}. Texas and California provide two distinct examples of U.S. market design, with different market structures, operational priorities, and levels of renewable integration \cite{texas_as, caiso_as}. Australia is included because its ancillary-service frameworks have evolved rapidly in response to high renewable penetration, system-strength challenges, and stability-related operational requirements \cite{au_as}.

The evidence base consists primarily of official system-operator publications, grid codes, market rules, regulatory documents, and technical reports, supplemented by relevant academic literature. Information from these sources was reviewed and organised around five comparison dimensions. First, service definition and classification encompass product scope, technical requirements, response times, and duration. Second, procurement and remuneration cover market-based and regulated frameworks, grid-code obligations, bilateral or contractual mechanisms, procurement timescales, pricing rules, and payment structures. Third, market organisation and coordination concern interactions between energy, balancing, and ancillary-service markets and operator roles. Fourth, technology participation and access cover renewable resources, BESS, demand-side resources, aggregated distributed resources, and other emerging providers. Fifth, recent reforms include faster response products, stability-related services, and dynamic, locational, or technology-neutral procurement.

Together, these dimensions provide a consistent basis for comparing how ancillary-service markets are structured and how they are adapting to changing system requirements. The objective is to examine how these systems respond to similar technical pressures through different market and regulatory arrangements and to identify transferable lessons and design challenges under increasing renewable penetration, rather than rank systems or prescribe a single optimal design.

\section{Comparative Analysis of Ancillary-Service Market Design}
\label{s4}

Rather than presenting separate country studies, this section applies the comparative dimensions defined in Section~\ref{s3} to the five service categories introduced in Section~\ref{s22}. This structure highlights cross-system differences in product granularity, procurement, remuneration, coordination, technology participation, and the translation of emerging system needs into explicit services.

\begin{figure}
\centering
\includegraphics[width=0.49\textwidth]{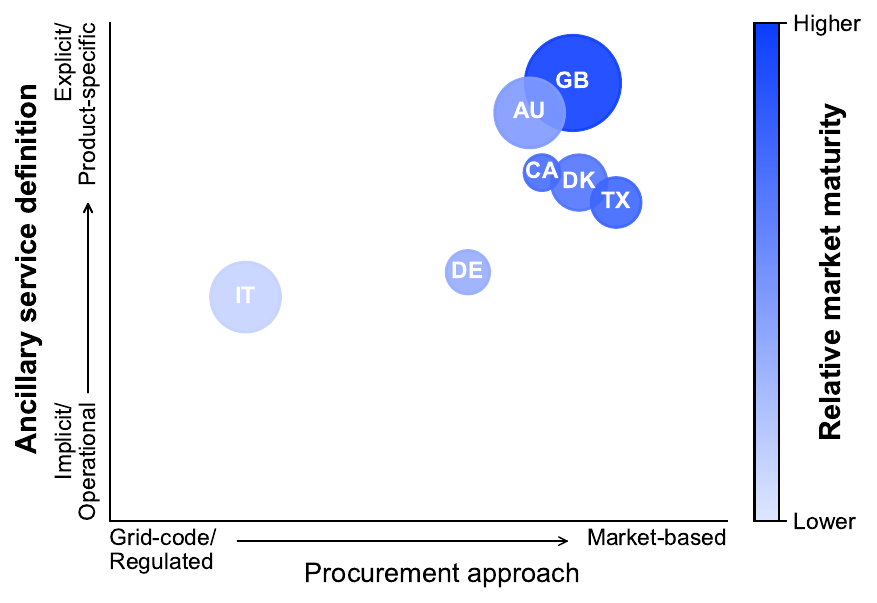}
\caption{Comparison of ancillary-service market design in seven electricity systems. Bubble size represents the diversity of ancillary-service products.}
\label{fig:as_design_2D}
\end{figure}

Figure~\ref{fig:as_design_2D} provides an aggregate qualitative comparison of ancillary-service market design among the reviewed electricity systems. The horizontal axis represents the procurement approach, ranging from grid-code, regulated, or operational arrangements to market-based procurement. The vertical axis reflects the degree to which ancillary services are defined explicitly as product-specific services rather than treated implicitly as operational capabilities. Bubble size indicates the diversity of ancillary-service products, while colour represents the relative maturity of each framework. The figure is intended as a conceptual comparison rather than a quantitative ranking. It highlights structural differences between market frameworks and provides a visual basis for the more detailed discussion developed in the following subsections.

\subsection{Service Design and Product Classification}

Most electricity systems require ancillary services to address similar operational needs, but they translate these needs into different product definitions, technical requirements, and performance obligations. These differences reflect trade-offs between standardisation, operational flexibility, market transparency, and product complexity \cite{11415609, onenet}. Although the services can be grouped into the five categories introduced in Section~\ref{s22}, the specific products within each category vary considerably across systems.

\subsubsection{Frequency Control and Reserve Services}

Frequency control and reserve services are the most extensively developed and explicitly defined category of ancillary services \cite{entsoe,SCHERER2013292,FERNANDEZMUNOZ2020109662}. Despite differences in terminology and implementation, all systems organise frequency control through a layered sequence of actions operating over different timeframes. Faster services arrest and contain frequency deviations, while slower services restore system balance, replace earlier responses, and replenish reserves for subsequent events. Across the reviewed systems, the main differences concern three aspects: the degree of standardisation in the service hierarchy, the granularity of response-speed categories, and the extent to which frequency services are integrated with broader market-clearing processes.

In Germany, Denmark, and Italy, frequency control follows a relatively standardised hierarchy centred on frequency containment reserve (FCR), automatic frequency restoration reserve (aFRR), and manual frequency restoration reserve (mFRR). These products follow broadly common activation principles and timescales: FCR provides the fastest containment response, while aFRR and mFRR progressively restore system balance over longer timeframes. A common feature of aFRR and mFRR is the separation between capacity and energy components\footnote{Capacity is procured mainly at the national or regional level, while activated balancing energy is exchanged via the platform for the international coordination of automatic frequency restoration and stable system operation (PICASSO) for aFRR and the Manually Activated Reserves Initiative (MARI) for mFRR.} \cite{ENTSOE2025MarketReview, germany_as, energinetasdenmarkconditions, terna_as_7page}. This structure supports harmonisation and cross-border coordination through European balancing platforms (i.e., PICASSO, MARI) \cite{eu_balancing,ENTSOE2025MarketReview,nordicafrr,nordicmfrr}.

\begin{figure*}[h!]
\centering
\includegraphics[width=0.95\textwidth]{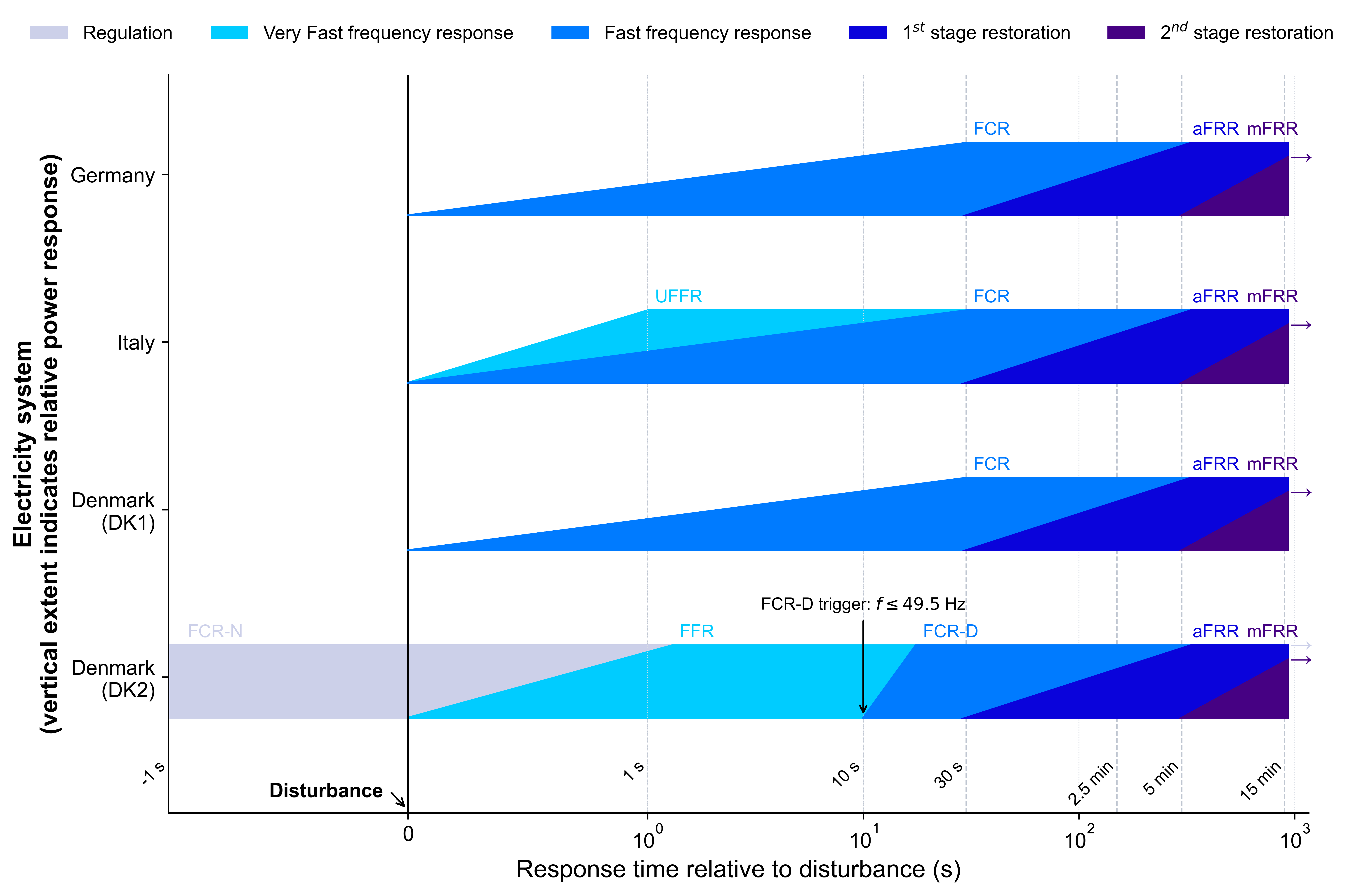}
\caption{Indicative comparison of response time and sustainment duration for selected European frequency-control products \cite{germany_as, energinetasdenmarkconditions, italy_grid_code}.}
\begin{minipage}{\textwidth}
\footnotesize
\textit{Note:} Negative times denote pre-fault or continuously active services. Diagonal segments indicate the time to full delivery, horizontal segments indicate sustained delivery, and arrows indicate continuation beyond the plotted time horizon. The vertical extent is schematic and does not represent directly comparable MW volumes across products or systems.
\end{minipage}
\label{fig:eu_frequency_timeline}
\end{figure*}

This standardised European approach is nevertheless complemented by system-specific products. In Denmark, Energinet procures additional fast-response products, particularly in the DK2 synchronous area, one of the two synchronous areas of the Danish power system (refer to Section~\ref{s43}), including fast frequency reserve (FFR) and distinct Nordic FCR variants, namely FCR for normal operation (FCR-N) and FCR for disturbance operation (FCR-D) \cite{nordicffr,nordicfcr}. In Italy, Terna includes ultra-fast frequency reserve (UFFR) as an additional fast frequency-response product \cite{italy_grid_code}. Replacement reserve (RR) extends the hierarchy beyond mFRR by providing a slower flexibility layer to restore reserve margins over longer timeframes. While RR is explicitly defined in some systems, such as Italy, it is less clearly separated in others, such as Germany and Denmark, where it is often integrated within broader balancing or reserve arrangements \cite{germany_as, energinetasdenmarkconditions, italy_grid_code}. These additions show that even relatively standardised European frameworks are becoming more granular as TSOs adapt to low-inertia and high-renewable operating conditions, while still maintaining a common frequency-control structure across the wider interconnected system. Figure~\ref{fig:eu_frequency_timeline} summarises the indicative response-time spectrum of these products.

\begin{figure*}
\centering
\includegraphics[width=0.95\textwidth]{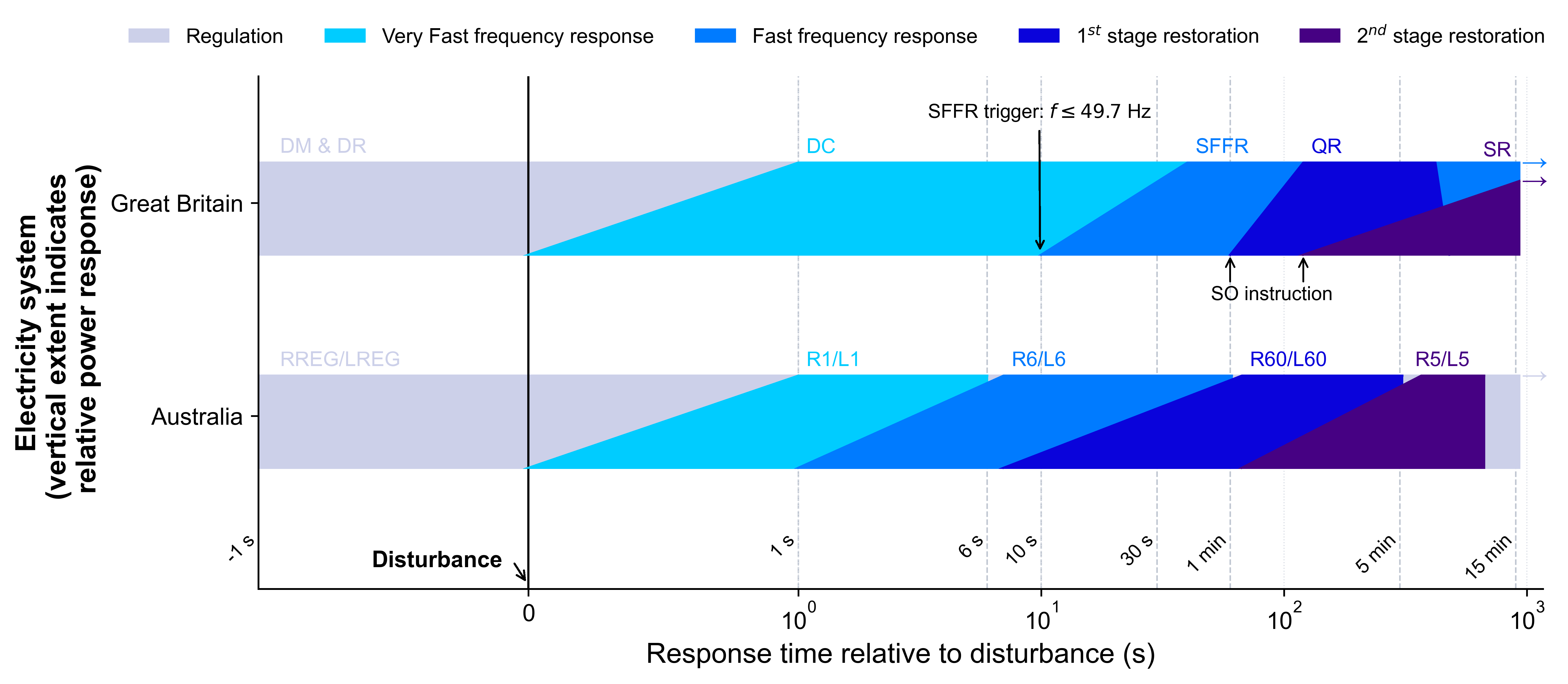}
\caption{Indicative comparison of response time and sustainment duration for frequency-control products in Great Britain and Australia \cite{neso_as, au_as}. The figure follows the same graphical conventions and interpretation as Figure~\ref{fig:eu_frequency_timeline}.}
\label{fig:uk_au_timescale}
\end{figure*}

\begin{figure*}
\centering
\includegraphics[width=0.95\textwidth]{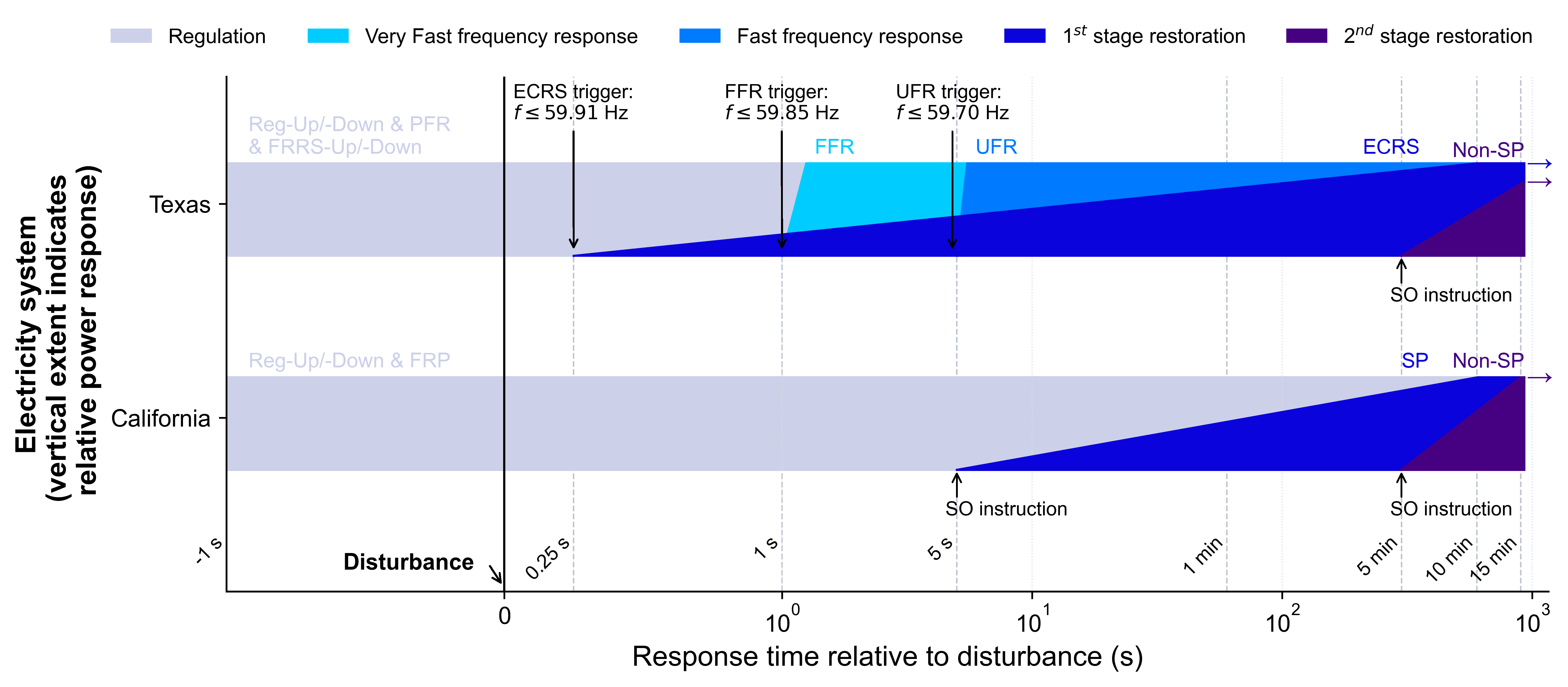}
\caption{Indicative comparison of response time and sustainment duration for frequency-control products in Texas and California \cite{texas_as, caiso_as}. The figure follows the same graphical conventions and interpretation as Figure~\ref{fig:eu_frequency_timeline}.}
\label{fig:tx_ca_timescale}
\end{figure*}

The timelines in Figures~\ref{fig:eu_frequency_timeline}--\ref{fig:tx_ca_timescale} provide an indicative product-by-product comparison of response time and sustainment duration rather than a single chronological activation sequence. Although slower services may operationally replace or replenish faster services, the intervals shown correspond to the specified delivery characteristics of each service considered individually. Negative times denote pre-fault or continuously active services, diagonal segments indicate the time required to reach full delivery, and horizontal segments indicate sustained delivery. Services extending beyond 15 minutes are truncated for visual clarity. For readability, upward/downward variants and symmetric/asymmetric requirements are not shown separately; further product-level characteristics are provided in Table~\ref{tab:summary_table} in Appendix~\ref{appendix1}.

Great Britain and Australia adopt more granular frequency-service frameworks, although they differ in how these products are integrated with market operations. In both systems, declining inertia and the need for very fast frequency support have encouraged the development of multiple products with distinct response speeds, durations, and operational roles \cite{8779818,ukfrfuture}. Compared with the European framework, these systems place greater emphasis on tailoring service definitions to domestic operating conditions rather than relying primarily on cross-border harmonisation.

In Great Britain, the framework has evolved from traditional products such as balancing reserve (BR), short-term operating reserve (STOR), fast reserve (FR), static firm frequency response (SFFR), and mandatory frequency response (MFR) towards a more differentiated set of frequency-response and reserve products. These include dynamic containment (DC), dynamic moderation (DM), dynamic regulation (DR), slow reserve (SR), which replaced STOR, and quick reserve (QR) \cite{neso_as}. These services differ not only in response speed, ranging from near-instantaneous response to several minutes, but also in operational role, including post-fault containment, continuous regulation, and reserve replacement. More recently, Great Britain has introduced co-optimised day-ahead procurement for selected response and reserve products through the Enduring Auction Capability, including the joint procurement of QR and selected dynamic response products \cite{ofgem_decision, neso_eac}. This is co-optimisation within ancillary-service procurement rather than full co-optimisation of energy and ancillary services.

Australia also uses a highly segmented frequency-control ancillary services (FCAS) framework, but it is more directly integrated with real-time energy dispatch. The FCAS structure separates contingency and regulation services across several response timescales. In particular, very fast contingency services, such as 1-second FCAS (R1/L1), complement the existing 6-second (R6/L6), 60-second (R60/L60), and 5-minute services (R5/L5). Dedicated regulation FCAS products, namely regulation raise and regulation lower (RREG/LREG), provide continuous frequency regulation under normal operating conditions \cite{au_as}. Unlike Great Britain, however, FCAS products are enabled and priced through real-time co-optimisation with energy dispatch. Australia therefore combines product granularity with an integrated dispatch-based design \cite{au_as}. These approaches are illustrated in Figure~\ref{fig:uk_au_timescale}, which visualises the different services and response speeds adopted in the Great Britain and Australian frameworks.

Texas and California represent a slightly different model, in which frequency-control and reserve services are closely integrated with energy-market operations through centralised market-clearing processes \cite{texas_as, caiso_as}. In these systems, ancillary services are generally co-optimised with energy in day-ahead and real-time markets, allowing resources to be allocated simultaneously across energy, reserve, and balancing requirements \cite{ercotas_study, caiso_bm}. This reduces the need for a highly disaggregated hierarchy of standalone frequency and reserve products because the optimisation process itself coordinates resource allocation across multiple operational needs.

In Texas, the framework combines traditional regulation and reserve services with several fast-response products \cite{texas_as}. Core services include regulation up/down (Reg-Up/-Down), responsive reserve service (RRS), non-spinning reserve (Non-SP), and emergency contingency reserve service (ECRS) \cite{ercotas_study}. RRS includes several very fast frequency-related capabilities, such as FFR, under-frequency response (UFR), and primary frequency response (PFR). Fast-responding regulation service up/down (FRRS-Up/-Down) also falls within the regulation category. Some of these services operate on sub-second or near-instantaneous timescales. Despite this increasing product differentiation, the Texas framework remains strongly linked to centralised co-optimisation \cite{texas_as, ercotas_study}.

California relies on a more compact set of ancillary-service products, primarily including Reg-Up/-Down, spinning reserve (SP)\footnote{Figure~\ref{fig:tx_ca_timescale} indicates 5 seconds activation time as an example. This service is activated upon SO instruction.}, and non-spinning reserve (Non-SP) \cite{caiso_as}. Similar to Texas, these services are co-optimised with energy in both day-ahead and real-time markets through integrated market-clearing processes \cite{caiso_fifth}. The Flexible Ramping Product (FRP) is also procured through real-time co-optimisation to ensure sufficient ramping capability for managing short-term variability and uncertainty associated with high renewable penetration \cite{caiso_bm}. As with the previous markets, Figure~\ref{fig:tx_ca_timescale} presents the response-time characteristics of the main frequency-control and reserve products in these two markets.

These contrasting approaches reveal a central trade-off in frequency-service design between standardisation, granularity, and market integration. First, standardised frameworks are well suited to interconnected systems because frequency is shared across the synchronous area, and common products can support coordination, liquidity, and cross-border balancing. In larger interconnected systems, higher aggregate inertia may also reduce the immediate need for very fast products compared with smaller or more electrically isolated systems. However, standardised frameworks may adapt more slowly because changes to product definitions, activation rules, or procurement arrangements often require agreement across multiple TSOs, regulators, and market participants. Second, granular product frameworks allow SOs to procure services that more closely match specific frequency dynamics, particularly under low-inertia conditions, and can facilitate participation by fast-acting inverter-based resources. However, greater product granularity can also increase market complexity, fragmentation, coordination requirements, and barriers for smaller or less flexible participants. Third, the level of market integration is also important. Co-optimised models can improve economic efficiency and operational coordination while reducing the need for multiple standalone products. However, the value of very fast or specialised capabilities must be represented adequately within the market-clearing model or addressed through complementary procurement arrangements. The evolution of frequency-control and reserve services reflects a broader shift towards faster response requirements, more differentiated service characteristics, and closer alignment between product design and system conditions. Although the reviewed systems retain a broadly similar layered structure, they implement it in different ways, reflecting differences in synchronous-area size, inertia levels, interconnection, operational risk, technology participation, and market-design objectives.

\subsubsection{Voltage and Reactive Power Support Services}

Voltage support and reactive power compensation differ fundamentally from frequency-control services because voltage is inherently location-dependent \cite{powercontrol, 917280}. While frequency is generally managed as a system-wide variable within a synchronous area, voltage varies by network node and is strongly affected by local demand and generation patterns, network topology, and reactive-power availability. This locational dependence makes voltage support difficult to standardise as a uniform product. Consequently, voltage and reactive power services are commonly delivered through grid-code obligations, SO instructions, bilateral contracts, or targeted tenders. In the systems considered, generators, inverter-based resources, and storage assets must provide reactive-power capability within specified technical limits. However, the definition, procurement, and remuneration of this capability vary considerably. The comparison below therefore examines three aspects: mandatory baseline capability, mechanisms for procuring additional support, and the degree to which procurement targets specific locations.

Great Britain adopts a hybrid approach, combining mandatory reactive power obligations with additional market-based and tendered arrangements \cite{neso_as, neso_stability}. Mandatory provision is implemented through the obligatory reactive power service (ORPS), which requires eligible generators to provide reactive power capability in accordance with grid-code requirements. Additional capability can be provided through the enhanced reactive power service (ERPS), which allows providers to offer reactive power beyond their mandatory obligations. Targeted voltage network services (VNS) extend this framework to specific locational needs. The Voltage 2026 tender, which represents the third phase of the Voltage Pathfinder programme, focuses on securing reactive power absorption capability in selected transmission regions \cite{neso_as}. The Pathfinder programme therefore provides an illustrative example of the transition from targeted procurement initiatives towards a broader reactive power framework covering long-, medium-, and short-term procurement horizons. These arrangements remain under development, and their interaction with the stability market is discussed in Section~\ref{section413} \cite{neso_stability}.

In Germany, Italy, and Denmark, mandatory technical requirements provide the common foundation, although the arrangements for securing additional support differ. In Germany, technical connection requirements establish baseline reactive-power capability, while market-based mechanisms may be used to procure additional support where needed \cite{germany_as,as_germany_paper,flex4fact}. Italy relies primarily on mandatory, location-specific requirements for voltage control and reactive power, as specified in its grid code \cite{italy_grid_code,flex4fact}. In Denmark, production units must be capable of providing voltage support and reactive-power compensation in accordance with grid-connection and generator requirements \cite{denmark_as,energinetasdenmarkconditions}. Germany therefore combines mandatory requirements with market-based procurement, whereas Italy and Denmark rely more heavily on technical obligations and network-specific operational requirements.

Texas and California also rely primarily on obligation-based provision. In Texas, the ERCOT Nodal Protocols establish reactive-power capability and voltage-control obligations for connected resources \cite{texas_as,texas_nodal_protocol}. California similarly treats voltage support as an operational requirement rather than as a product within the co-optimised ancillary-service market. If needed, additional capability may be secured through system-operator instructions or contractual arrangements \cite{caiso_as,caiso_bm}. Thus, in both systems, voltage support remains largely outside the centralised co-optimisation of energy and frequency-related ancillary services.

Australia differs in its use of a dedicated network-support framework. Voltage and reactive power needs are managed through network support and control ancillary services (NSCAS), which are non-market services used when standard dispatch mechanisms are insufficient to maintain secure network operation. Within this framework, reliability and security ancillary services (RSAS) provide additional operational flexibility to maintain voltage within acceptable limits, while market benefit ancillary services (MBAS) address voltage-related network limitations through technical and operational solutions that enhance network capability \cite{au_as}. This approach reflects the inherently locational nature of voltage management and the need for targeted solutions where local network constraints or system-security issues arise.

Overall, a mandatory or regulated technical capability remains the common foundation in the reviewed systems. The principal difference lies in how additional, location-specific needs are secured. Great Britain and Germany supplement baseline obligations with market-based procurement mechanisms. California and Australia rely primarily on contractual or non-market arrangements for additional capability, whereas Italy, Denmark, and Texas rely more heavily on regulated obligation-based requirements and system-operator actions. Unlike frequency-service reforms, which have largely focused on response speed, product granularity, and integration with balancing markets, developments in voltage and reactive power support centre on locational targeting, procurement horizons, and the boundary between mandatory capability and remunerated additional provision. This distinction is particularly relevant in high-renewable systems, where IBRs alongside synchronous condensers, shunt reactors, and other dynamic or static devices, are increasingly expected to contribute to voltage control and reactive power management. Table~\ref{tab:voltage_reactive_comparison} consolidates the comparison by summarising baseline provision, arrangements for securing additional support, and the locational implementation of voltage and reactive power services.

\begin{table*}[t]
\centering
\begin{threeparttable}
\caption{Comparison of voltage and reactive power support arrangements across the reviewed electricity systems.}
\label{tab:voltage_reactive_comparison}
\renewcommand{\arraystretch}{1.7}
\begin{tabular}{
p{0.12\textwidth}
p{0.25\textwidth}
p{0.29\textwidth}
p{0.25\textwidth}}
\hline
\textbf{System} &
\textbf{Baseline provision} &
\textbf{Additional procurement} &
\textbf{Locational implementation} \\
\hline

Great Britain \cite{neso_as, neso_stability} &
Mandatory provision through ORPS &
ERPS and targeted VNS procurement &
Explicitly locational, including longer-term regional tenders
\\

Germany \cite{germany_as, as_germany_paper, flex4fact} &
Mandatory capability under technical connection requirements &
Market-based procurement where additional reactive power is required &
Applied according to network-specific requirements
\\

Italy \cite{italy_grid_code, flex4fact} &
Mandatory voltage-control and reactive power capability &
Primarily secured through regulated and operational arrangements &
Locational requirements specified through the grid code
\\

Denmark \cite{denmark_as, energinetasdenmarkconditions} &
Mandatory technical capability for connected production units &
Additional needs addressed through system-operator arrangements &
Determined by grid-connection and local network requirements
\\

Texas \cite{texas_as, texas_nodal_protocol}  &
Mandatory reactive power capability and voltage-control obligations (ERCOT protocols) &
Additional support directed through ERCOT operational arrangements &
Managed according to nodal and local voltage requirements
\\

California \cite{caiso_as, caiso_bm}  &
Operational voltage-support obligations &
Additional support through system-operator instructions or contracts &
Managed outside the co-optimised ancillary-service market
\\

Australia \cite{au_as}  &
Technical capability supported by the NSCAS framework &
RSAS and MBAS procured through non-market or contractual arrangements &
Targeted towards identified network-security and capability needs
\\

\hline
\end{tabular}
\end{threeparttable}
\end{table*}

\subsubsection{System Stability Services}
\label{section413}

System stability services constitute a broad and evolving category of ancillary services. They are less standardised and less consistently defined than frequency-control, voltage-support, and reactive-power services \cite{9796617}. Their main purpose is to address the operational challenges associated with declining synchronous generation, low inertia, reduced system strength, and interactions among IBRs, thereby supporting secure operation under normal conditions and following disturbances \cite{Ahmed2023,pierluigistabilitty,8450880}. These services include inertia provision, system-strength support, DVS, oscillation damping, and short-circuit current contribution \cite{entsoe2021,DREIDY2017144,9796617}. Although system strength directly affects local voltage behaviour, this review treats it as a stability requirement because it concerns voltage-waveform robustness, inverter interactions, fault-level adequacy, and post-disturbance behaviour rather than voltage-magnitude and reactive-power control alone. Recent developments have focused particularly on inertia and system strength as IBR penetration grows and synchronous generation declines \cite{entsoeinertia}. Compared to other ancillary-service categories, stability services remain at an earlier stage of market and regulatory development, with substantial variation in how these capabilities are defined, procured, and integrated into system operation.

Three broad approaches emerge from the comparison. First, stability capabilities may be procured through explicit or developing market-based arrangements. Great Britain has established explicit stability-service procurement, while Germany is moving towards market-based inertia procurement. Second, stability needs may be addressed through structured non-market arrangements, including network-support agreements and transitional service frameworks, as in Australia. Third, stability may be managed primarily through technical requirements, operational measures, and the deployment of advanced grid-support technologies rather than dedicated ancillary-service products. This approach is more evident in Italy, Denmark, Texas, and California.

Great Britain represents one of the most straightforward examples of system stability service development. Stability network services (SNS) were introduced through the three-phase Stability Pathfinder programme \cite{neso_as, neso_stability}. These initiatives procured specific capabilities related to inertia, SCL, and DVS. Following the Pathfinder phases, Great Britain has been developing a broader stability-market framework with long-term, medium-term, and short-term procurement horizons, similar to the reactive power market discussed above \cite{neso_stability}. The medium-term market has progressed through several procurement rounds, while the long-term and short-term arrangements remain under development \cite{neso_as, neso_stability}. The long-term framework is intended to support joint procurement under the stability and reactive power markets, using a bundled tender structure to secure multiple capabilities for future system needs from 2029 onwards \cite{neso_as}. This represents a transition from targeted Pathfinder procurement towards a more enduring and integrated stability-service framework.

In continental European systems, stability-related capabilities are generally less clearly defined as dedicated market products. In Germany, the TSOs are developing inertia services through market-based mechanisms, including product categories differentiated by availability and directional characteristics \cite{germany_as,as_germany_paper,germany_inertia}. However, explicit stability procurement remains limited. In Italy, Terna mainly addresses inertia, SCL, and oscillation management through mandatory, location-specific grid-code requirements \cite{italy_grid_code}. Terna also distinguishes between mechanical and synthetic inertia contributions, reflecting the potential role of appropriately controlled IBRs in supporting system stability. In Denmark, Energinet has not yet procured a dedicated inertia product. FFR, particularly in the DK2 synchronous area, supports operation under low-inertia conditions by providing a rapid active-power response following disturbances \cite{denmark_as,energinetasdenmarkconditions}. However, it does not provide or replace physical inertia; rather, it complements inertia by helping to arrest frequency deviations after a disturbance. Current developments also consider the potential contribution of synthetic inertia from appropriately controlled IBRs.

In Texas and California, there are currently no ancillary-service products specifically dedicated to system stability. In Texas, inertia adequacy is addressed mainly through monitoring and operational intervention rather than market-based procurement. Additional synchronous generation may be committed when system inertia approaches critical operating thresholds \cite{texas_as}. In parallel, grid-forming technologies and synchronous condensers are being considered, while enhanced technical requirements for IBRs are being introduced to strengthen system stability \cite{ercot_grid_insights}. California has similarly emphasised technical requirements and the deployment of resources equipped with grid-forming or other advanced inverter controls, including appropriately configured BESS, rather than establishing dedicated stability products \cite{caiso_as}. Such resources can provide fast dynamic support and improve stability performance when the required control capabilities are enabled. More broadly, California's ancillary-service framework has been progressively adapted to accommodate fast-acting, non-traditional resources \cite{caiso_fifth}. However, the participation of these resources in existing ancillary-service markets should be distinguished from the explicit procurement of inertia or system strength.

In Australia, system stability requirements are managed mainly through broader network-support and transitional-service frameworks rather than through a single dedicated product \cite{8779818}. Stability-related needs can be met through the NSCAS framework when standard market and operational mechanisms are not enough. Within this framework, RSAS address security needs associated with system strength and inertia, while MBAS may address stability-related network limitations when these constrain transfer capability or network utilisation \cite{au_as}. In addition, Transitional Services provide a further mechanism for filling emerging capability gaps as the system moves towards operation with low or zero levels of synchronous generation \cite{au_as}. These arrangements can support the deployment and testing of grid-forming assets, synchronous condensers, and other stability-support technologies.

System stability services remain more system-specific and less harmonised than frequency-control services because their value depends strongly on local network conditions, the generation mix, system strength, and the availability of synchronous resources. As IBR penetration grows, capabilities such as inertia, fault-current contribution, dynamic voltage support, and system-strength support must be defined and procured more explicitly. Grid-forming resources, synchronous condensers, and advanced control capabilities offer potential technical solutions, but the corresponding market and regulatory arrangements remain uneven. As synchronous generation declines, these services are likely to become increasingly important for the secure and reliable operation of low-carbon power systems. Table~\ref{tab:stability_services} maps the main system stability capabilities and their corresponding procurement, regulatory, or operational arrangements.

\begin{table*}[t]
\centering
\caption{System stability service arrangements across the reviewed electricity systems.}
\label{tab:stability_services}
\renewcommand{\arraystretch}{1.7}
\begin{tabular}{
p{0.20\textwidth}
p{0.27\textwidth}
p{0.45\textwidth}}
\hline
\textbf{System} &
\textbf{Main stability focus} &
\textbf{Implementation approach} \\
\hline

Great Britain
\cite{neso_as, neso_stability}
&
Inertia, SCL, and DVS
&
Dedicated SNS procurement and emerging multi-horizon stability markets
\\

Germany
\cite{germany_as, as_germany_paper, germany_inertia}
&
Inertia
&
Emerging market-based procurement alongside technical requirements
\\

Italy
\cite{italy_grid_code}
&
Inertia, SCL, and oscillation management
&
Mandatory and locational grid-code requirements
\\

Denmark
\cite{denmark_as, energinetasdenmarkconditions}
&
Low-inertia operation
&
Technical requirements and FFR, no dedicated inertia product
\\

Texas
\cite{texas_as, ercot_grid_insights}
&
Inertia adequacy and dynamic performance
&
Operational monitoring, synchronous commitment, and technical requirements
\\

California
\cite{caiso_as, caiso_fifth}
&
Dynamic stability and grid-forming capability
&
Technical requirements and advanced IBR deployment, no dedicated stability product
\\

Australia
\cite{8779818, au_as}
&
System strength, inertia, and low-synchronous operation
&
NSCAS and Transitional Services
\\

\hline
\end{tabular}
\end{table*}

\subsubsection{Restoration Services}

Restoration services are generally delivered through regulated arrangements, contracts, targeted tenders, or operational processes rather than through centrally traded ancillary-service markets. Across the selected electricity systems, restoration capability is typically secured in advance from eligible resources to support system recovery following a partial or complete blackout. In Great Britain, this capability is secured through long-term arrangements and is increasingly coordinated with broader system-support requirements \cite{neso_stability}. Germany and Denmark also rely on tenders or long-term bilateral arrangements \cite{germany_as, energinetasdenmarkconditions}, whereas Italy implements restoration capability mainly on mandatory, location-specific grid-code requirements as part of its power-system restoration and recovery process \cite{italy_grid_code}. Outside Europe, Texas uses bilateral agreements with eligible providers \cite{texas_as}, while California secures restoration capability through administrative or contractual arrangements \cite{caiso_bm}. Australia uses a dedicated system restart ancillary service (SRAS), which is procured contractually through open competitive tenders or direct requests for offers \cite{au_as}. Overall, restoration services are treated primarily as reliability and resilience capabilities rather than as continuously traded products, and their procurement remains highly system-specific because it must reflect each system’s restoration strategy, network characteristics, eligible resources, and operational requirements.

\subsubsection{Congestion Management Services}

Congestion management is highly location-dependent, and its implementation varies substantially among market designs. In some electricity systems, network constraints are managed through redispatch, regulation, or targeted network-support services. In others, congestion management is embedded directly within nodal market clearing and is therefore not defined as a separate ancillary-service product.

In Germany, Denmark, and Italy, the relevant TSOs mainly manage network constraints through redispatch, regulation, or location-specific modulation mechanisms. In Germany, the TSOs redispatch generation and storage resources to relieve network bottlenecks \cite{germany_as}. Denmark uses upward and downward regulation to address temporary transmission constraints \cite{energinetasdenmarkconditions}, while in Italy, ``extraordinary modulation`` services adjust active power when standard balancing and redispatch actions are insufficient \cite{italy_grid_code}.

Texas and California rely primarily on nodal market dispatch. In Texas, congestion is managed mainly through security-constrained economic dispatch (SCED), which redispatches generation in real time to maintain power flows within network limits \cite{texas_as}. In California, congestion is managed through day-ahead and real-time market clearing using locational marginal pricing (LMP) \cite{caiso_bm,caiso_fifth}. In both systems, congestion management is integrated into the energy-market dispatch process rather than procured as a separate ancillary-service product.

Great Britain and Australia combine operational dispatch with targeted network-support arrangements. In Great Britain, services such as the constraint management intertrip service (CMIS), MW Dispatch, and the transmission constraint management service (TCMS) are used to address specific network constraints and reduce the cost of corrective actions \cite{neso_as}. In Australia, RSAS and MBAS within the broader NSCAS framework can be used to address network-security or transfer-capability limitations beyond the scope of routine market dispatch \cite{au_as}. These arrangements enable targeted intervention when specific constraints cannot be managed adequately through standard dispatch alone.

The comparison shows that the main distinction lies in whether congestion management is implemented through targeted services, operational redispatch, or nodal market clearing. Because its value depends on the location, severity, and duration of specific network constraints, congestion management is less suited to uniform product standardisation. Its design is therefore closely shaped by network topology, market architecture, and system-operator practices.

\paragraph{Cross-category synthesis:} Figure~\ref{fig:cross_category_grid} compares the dominant implementation approaches across the five service categories and seven systems. Rows represent service categories, columns represent electricity systems, and circle styles distinguish implementation structures, as defined in the caption, rather than market maturity or performance. Frequency control and reserve services are the most consistently formalised through explicit products or integrated market mechanisms. Implementation approaches for voltage and reactive-power support and system stability vary more widely, reflecting their dependence on local network conditions, technical requirements, and system-specific procurement needs. Restoration is generally secured through contractual, tendered, or regulated arrangements, while congestion management ranges from operational redispatch and targeted services to nodal market clearing. The degree of standardisation therefore decreases as services become more locational, network-dependent, and closely tied to system-specific operational conditions.

\newsavebox{\asopenbox}
\newsavebox{\ashalfbox}
\newsavebox{\asfullbox}

\savebox{\asopenbox}{%
  \tikz[baseline=-0.45ex]{\draw[line width=0.7pt] (0,0) circle (0.36em);}}
\savebox{\ashalfbox}{%
  \tikz[baseline=-0.45ex]{%
    \fill (0,0) -- ++(90:0.36em)
      arc[start angle=90, end angle=-90, radius=0.36em] -- cycle;
    \draw[line width=0.7pt] (0,0) circle (0.36em);}}
\savebox{\asfullbox}{%
  \tikz[baseline=-0.45ex]{\fill (0,0) circle (0.36em);}}

\DeclareRobustCommand{\asopen}{\usebox{\asopenbox}}
\DeclareRobustCommand{\ashalf}{\usebox{\ashalfbox}}
\DeclareRobustCommand{\asfull}{\usebox{\asfullbox}}

\begin{figure*}[t]
\centering
\includegraphics[width=0.90\textwidth]
{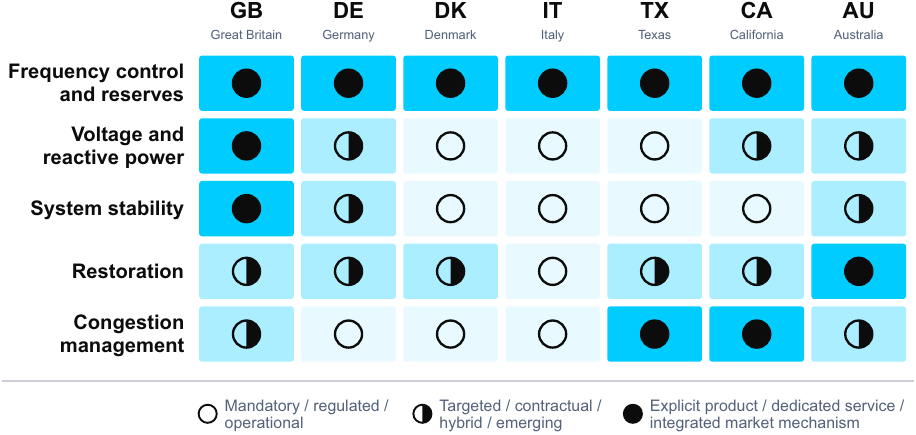}
\caption{Cross-category comparison of ancillary-service implementation in seven electricity systems. Rows represent service categories, and columns represent electricity systems. Open circles \asopen{} indicate mandatory, regulated, or operational provision; half-filled circles \ashalf{} indicate targeted, contractual, hybrid, or emerging arrangements; and filled circles \asfull{} indicate explicit products, dedicated services, or integrated market mechanisms. These classifications describe implementation structures and do not constitute a ranking of market maturity, effectiveness, or performance.}
\label{fig:cross_category_grid}
\end{figure*}

\subsection{Procurement and Remuneration Mechanisms}
\label{s42}

Building on the preceding comparison, this subsection examines procurement horizons, price formation, and payment structures. These are related but distinct aspects of ancillary-service design: procurement determines how capability is secured, pricing rules determine remuneration rates, and payment structures specify which contributions are compensated. A detailed system- and product-level comparison is provided in Table~\ref{tab:summary_table} in Appendix~\ref{appendix1}.

Procurement may take the form of recurring market-based auctions, targeted tenders, or bilateral contracts. These approaches differ in how prices are determined, how much operational control the SO retains, and what investment incentives they provide. Market-based auctions can support transparent short-term price formation and reveal the opportunity costs of service provision, but they require sufficiently liquid markets, clearly defined products, and clear qualification rules. Tenders and bilateral contracts can secure specialised or location-specific capability over longer horizons, although their price signals may be less transparent and their terms less responsive to changing system conditions \cite{4077136,competition_pollitt,Rancilio2022}.

Procurement horizons reflect both the frequency with which requirements change and the investment commitments needed to meet them. Short-term procurement allows volumes to be adjusted as operating conditions evolve and is particularly common for frequency-control and reserve services. Longer-term procurement provides advance assurance of availability where provision depends on specialised assets, location, or new investment. For instance, Great Britain’s developing procurement horizons for voltage and stability services illustrate how different timescales can be combined within a broader framework \cite{neso_as,neso_stability}. Alongside these arrangements, grid codes, connection requirements, and system-operator instructions establish obligations for capability or delivery; these rely more heavily on administrative rules and may provide weaker incentives for investment and participation \cite{4077136,competition_pollitt,Rancilio2022}.

Remuneration arrangements generally reflect how capability is secured, although a given procurement mechanism does not necessarily imply a particular pricing rule. Standardised services procured through recurring auctions or integrated market clearing are commonly remunerated at market-clearing prices. This is also known as ``pay-as-clear'' settlement and is applied to selected European balancing products, Great Britain’s dynamic frequency-response auctions, and Australia’s FCAS markets \cite{ENTSOE2025MarketReview,neso_as,germany_as,au_as,energinetasdenmarkconditions,italy_grid_code}. Texas and California similarly use market-clearing prices for their principal frequency-control and reserve products \cite{ercotas_study,caiso_bm}. Such pricing is most readily applied where service requirements are clearly specified and offers from different providers can be compared. Settlement rules may nevertheless differ between capacity and energy components within the same service.

Services secured through targeted tenders or bilateral contracts more commonly use pay-as-bid or contract-specific remuneration. These arrangements allow payment terms to reflect the required capability, location, and availability period. Examples include pay-as-bid remuneration for restoration services in Germany and Denmark and contract-specific payments for Australian SRAS and NSCAS \cite{germany_as,energinetasdenmarkconditions,au_as}. In addition to determining remuneration rates, these agreements specify what providers are paid for. Providers may receive payments for availability, utilisation, or a combination of these, with some agreements also compensating testing or other specified activities. The distinction between pricing rules and payment components is therefore important: pricing rules determine remuneration rates, whereas payment components specify which obligations and activities are compensated.

Fixed-price or regulated remuneration is another approach, commonly associated with mandatory baseline capabilities and administratively specified services, although some mandatory capabilities receive no separate payment. For example, Great Britain’s ORPS combines mandatory reactive-power capability with fixed-price remuneration, whereas inertia provision in Italy generally receives no direct payment \cite{neso_as,italy_grid_code}. Fixed prices are also used outside mandatory provision, as illustrated by Germany’s market-based inertia arrangements, where payments depend on availability and product characteristics \cite{germany_as,germany_inertia,as_germany_paper,flex4fact}. These examples show that the obligation to provide a capability and the method used to remunerate it are distinct aspects of service design.

Procurement and remuneration arrangements reflect both the characteristics of the service and the institutional framework in which it is provided. Procurement horizons, pricing rules, and payment components jointly shape providers’ incentives to maintain availability, deliver services, and invest in capability. Effective design therefore requires these elements to be aligned, balancing responsiveness to changing operating conditions with sufficient revenue predictability to support investment, particularly in specialised or location-specific assets.

\subsection{Technology Participation and Market Access}
\label{s44}

Whereas Section~\ref{s42} examined how ancillary services are procured and remunerated, this subsection considers technology participation and the conditions governing market access. Participation in frequency-control and reserve services is increasingly determined by demonstrated performance rather than by technology type alone, broadening opportunities for providers such as BESS and demand-side resources \cite{neso_as,germany_as,texas_as,caiso_as,au_as,energinetasdenmarkconditions,italy_grid_code}. These technologies are particularly well suited to fast-response products because of their rapid controllability \cite{neso_as,texas_as,au_as,energinetasdenmarkconditions}. Eligibility nevertheless depends on meeting product-specific criteria related to response speed, sustainment duration, availability, telemetry, and prequalification.

Access to voltage, reactive-power support and system stability services remains more selective because eligibility depends on connection point, physical capability, and demonstrated performance under relevant network conditions \cite{germany_as,au_as,energinetasdenmarkconditions,italy_grid_code,caiso_bm,neso_stability}. Technologies such as synchronous condensers and appropriately controlled IBRs may participate, but their suitability must be assessed against the specific capability being procured. For IBRs, access may require grid-forming or other advanced control functions, supported by technical validation. Participation therefore depends on the resource's configuration and location as well as its technology type.

Restoration and congestion-management services also impose system-specific qualification and locational requirements. Appropriately configured storage, for example, may qualify for restoration provision subject to testing and compliance with the relevant restoration plan. DER aggregations may participate in congestion management where their location, controllability, and verified network impact meet operational requirements. Access in these categories therefore depends on both demonstrated capability and suitability for the particular network need.

Table~\ref{tab:technology_participation} summarises the main participating technologies and access conditions for the five ancillary-service categories. The comparison indicates that technology-neutral participation is most developed for frequency-control and reserve services, while access to the other categories depends more strongly on location and specialised technical capabilities. Expanding participation requires proportionate prequalification procedures, telemetry requirements, aggregation rules, and performance-verification methods that recognise the capabilities of emerging providers while maintaining system-security standards.

\begin{table*}[t]
\centering
\caption{Technology participation and access conditions across ancillary-service categories.}
\label{tab:technology_participation}
\renewcommand{\arraystretch}{1.7}
\small

\begin{tabular}{
p{0.2\textwidth}
p{0.38\textwidth}
p{0.34\textwidth}}
\hline
\textbf{Service category} &
\textbf{Main participating technologies} &
\textbf{Main access conditions} \\
\hline

Frequency control and reserves
\cite{neso_as, germany_as, texas_as, caiso_as, au_as, energinetasdenmarkconditions, italy_grid_code, fcrnfcrdenerginet}
&
Synchronous generators, BESS, demand-side resources, DER aggregations, flexible loads, and controllable renewable generation
&
Response speed, sustainment, controllability, telemetry, availability, and prequalification
\\

Voltage and reactive power support
\cite{germany_as, au_as, energinetasdenmarkconditions, italy_grid_code, caiso_bm, neso_stability, fcrnfcrdenerginet}
&
Synchronous generators, IBRs, STATCOMs, SVCs, capacitors, reactors, and synchronous condensers
&
Reactive power range, voltage-control capability, location, and grid-code compliance
\\

System stability
\cite{neso_stability, germany_inertia, italy_grid_code, ercot_grid_insights, caiso_as, au_as, energinetasdenmarkconditions, fcrnfcrdenerginet}
&
Synchronous generators and condensers, grid-forming IBRs, appropriately configured BESS, and HVDC systems
&
Required stability capability, location, dynamic performance, and technical validation
\\

Restoration
\cite{neso_stability, germany_as, energinetasdenmarkconditions, italy_grid_code, texas_as, caiso_bm, au_as, fcrnfcrdenerginet}
&
Black-start generators, appropriately configured storage, HVDC systems, and qualified DERs
&
Self-start or re-energisation capability, location, testing, and restoration-plan compliance
\\

Congestion management
\cite{neso_as, germany_as, denmark_as, italy_grid_code, texas_as, caiso_bm, au_as, fcrnfcrdenerginet}
&
Generators, storage, demand-side resources, DER aggregations, intertrip providers, and network-support assets
&
Locational suitability, controllability, network impact, operational availability, and qualification
\\

\hline
\end{tabular}
\end{table*}

\subsection{Market Structure and Coordination}
\label{s43}

Having considered procurement, remuneration, and market access, this subsection examines how ancillary services interact with wider electricity-market and system-operation arrangements. Coordination between ancillary services, energy trading, balancing actions, congestion management, and real-time operation is shaped by market coupling, pricing design, dispatch arrangements, balancing responsibility, and network representation. The reviewed electricity systems exhibit three distinct coordination patterns. Germany, Italy, and Denmark participate in common European energy and balancing platforms while retaining national or zonal mechanisms for system-specific requirements. Great Britain operates separate but interacting balancing and ancillary-service frameworks. Texas, California, and Australia rely more strongly on centralised dispatch and, particularly for frequency-related services, integrated energy and ancillary-service clearing. These patterns are not absolute, since locational and system-security services frequently remain outside common market-clearing arrangements.

European electricity and balancing markets are increasingly coordinated through common platforms. Day-ahead and intraday energy trading are organised through single day-ahead coupling (SDAC) and single intraday coupling (SIDC), while balancing-energy exchange is progressively integrated through platforms such as PICASSO for aFRR and MARI for mFRR \cite{entsoe}. Additional arrangements support imbalance netting, replacement-reserve exchange, frequency-containment cooperation, and the management of cross-zonal balancing capacity \cite{ENTSOE2025MarketReview, eu_balancing}. These mechanisms promote harmonisation and cross-border balancing efficiency, but they do not remove the need for national or zonal intervention. Internal congestion, voltage support, system stability, restoration, and other locational requirements remain primarily coordinated by the relevant TSOs.

The interaction between European coordination and national market design differs across Germany, Italy, and Denmark. Germany operates as a single bidding zone, supporting uniform wholesale price formation but leaving internal transmission constraints outside the energy-price signal. The geographical separation between substantial renewable generation in northern Germany and major demand centres in the south therefore creates a continuing need for redispatch and other congestion-management measures \cite{germany_as, flex4fact}. Italy also participates in European market coupling, but operates with seven bidding zones and zonal electricity prices. After the energy markets clear, Terna uses the dispatching services market, \textit{Mercato dei Servizi di Dispacciamento} (MSD), including its ex-ante scheduling stage and the balancing market, \textit{Mercato del Bilanciamento} (MB), to procure reserves, manage congestion, and address technical constraints \cite{italy_grid_code, flex4fact, terna_as}. More recently, under the TIDE reform, this framework has transitioned to the Balancing and Redispatching Market, \textit{Mercato di Bilanciamento e Ridispacciamento} (MBR), which integrates the scheduling and balancing functions of the former MSD and MB with participation in the European balancing platforms for standard balancing-energy products. Italy therefore combines European integration with zonal pricing and dedicated national dispatching arrangements.

Denmark participates in both European and Nordic coordination frameworks. Its electricity system is divided into DK1 and DK2: DK1 is synchronously connected to continental Europe, whereas DK2 belongs to the Nordic synchronous area \cite{denmark_as, energinetasdenmarkconditions}. The two bidding zones are connected through an HVDC link, and the available transfer capacity between them affects energy prices, reserve procurement, and the design and dimensioning of ancillary services. Denmark therefore illustrates how participation in common market platforms can coexist with different synchronous-area requirements within the same national electricity system.

Great Britain follows a different coordination structure based on separate but interacting market and operational mechanisms. Most energy is traded through forward markets, power exchanges, or bilateral contracts before delivery, while NESO uses the Balancing Mechanism (BM), ancillary-service arrangements, and short-term energy trading closer to real time to maintain system balance and security \cite{neso_as}. Ancillary services are not jointly cleared with energy through a single integrated market platform. Instead, they are procured through service-specific auctions, tenders, contracts, and mandatory arrangements that interact with the BM and wider balancing requirements \cite{neso_stability}. This structure allows targeted products to be developed for particular operational needs, but requires coordination between contracted services, balancing actions, and real-time dispatch.

Texas and California rely more directly on nodal pricing and centralised market-clearing processes. In Texas, ERCOT coordinates energy, ancillary services, and network constraints through day-ahead and real-time market processes, with real-time dispatch undertaken at five-minute intervals \cite{texas_as}. Energy and ancillary services are co-optimised in the day-ahead market, while real-time security-constrained economic dispatch updates resource schedules as operating conditions change \cite{ercotas_study, ercot_grid_insights}. California similarly coordinates energy, ancillary services, and congestion through nodal day-ahead and real-time markets \cite{caiso_as, caiso_bm}. Energy and ancillary services are co-optimised, and real-time processes progressively update dispatch closer to delivery \cite{caiso_fifth}. Unlike ERCOT, however, California is more strongly interconnected with neighbouring balancing areas, making imports, exports, transfer capability, and deliverability constraints more prominent in market coordination.

Australia also uses centralised dispatch, but differs from Texas and California in its pricing structure and market timeline. The National Electricity Market (NEM) applies regional rather than nodal pricing and is organised primarily around five-minute real-time dispatch rather than a formal day-ahead market \cite{au_as}. Energy and FCAS are co-optimised within this process, whereas stability, restoration, and network-support requirements are coordinated through separate non-market or contractual arrangements. Australia therefore combines integrated real-time dispatch of energy and FCAS with distinct frameworks for services that are more locational or system-specific.

The comparison shows that market architecture determines whether coordination is achieved through common cross-border platforms, interaction among separate service-specific mechanisms, or integrated centralised dispatch. European coordination supports harmonisation and cross-border balancing but continues to depend on national and zonal actions for network-specific requirements. Great Britain gains flexibility from separate service frameworks but must coordinate them carefully with balancing and real-time operation. Texas, California, and Australia integrate energy and frequency-related ancillary services more directly, although their network representation, pricing structures, and market timelines differ. None of these arrangements eliminates the need for separate mechanisms where service value depends strongly on location, network conditions, or specialised system-security requirements. As renewable penetration increases, effective coordination across energy, balancing, network, and ancillary-service arrangements becomes increasingly important.

\section{Challenges, Trends, and Market Design Implications}
\label{s5}

\subsection{Key Challenges in Ancillary-Service Markets}
The comparison in Section~\ref{s4} identifies four recurring challenges as electricity systems accommodate higher shares of renewable generation and in broader context, inverter-based resources. These concern the harmonisation of service frameworks, the valuation of flexibility and stability capabilities, participation by emerging resources, and coordination across market and operational mechanisms.

\subsubsection{Limited harmonisation, particularly for non-frequency services}
European mechanisms such as SDAC, SIDC, PICASSO, MARI, IGCC, and FCR Cooperation have strengthened cross-border coordination for energy trading and balancing services. However, harmonisation remains uneven across ancillary-service categories. Frequency-control and reserve products increasingly follow common definitions and balancing arrangements, whereas voltage support, system stability, restoration, and congestion management remain strongly system- and location-specific in their technical requirements, procurement rules, and remuneration mechanisms \cite{ENTSOE2025MarketReview, eu_balancing}. Similar challenges can arise in other interconnected systems where institutional, geographical, technical, or commercial boundaries limit the extent to which services can be standardised across regions.

\subsubsection{Incomplete valuation of flexibility and stability capabilities} 
Standardised frequency-control and reserve services can generally be priced through recurring auctions or integrated market-clearing mechanisms because their performance requirements are measurable and providers are comparatively substitutable. By contrast, inertia, SCL, DVS, system-strength support, voltage support, reactive power capability, and restoration are more difficult to value through uniform prices because their contribution depends on location, system conditions, and the availability of alternative resources. They are therefore frequently remunerated through tenders, contracts, regulated payments, or technical obligations, which may provide less transparent investment signals. The development of explicit stability procurement in Great Britain and emerging inertia arrangements in Germany illustrates a gradual move towards more explicit valuation of these capabilities. The remaining challenge is to reflect both their system-wide and locational value without imposing market structures that are poorly suited to their physical characteristics.

\subsubsection{Participation barriers for emerging resources} 
IBRs, demand-side resources, DER aggregations, storage, and flexible loads are increasingly capable of providing ancillary services, particularly fast frequency response and flexibility-oriented products. IBRs do not provide physical synchronous inertia, but appropriately controlled resources can provide fast active-power response, grid-forming functionality, and other dynamic support capabilities. Access nevertheless depends on prequalification rules, minimum size thresholds, telemetry and control requirements, response speed, sustainment duration, state-of-charge constraints, and locational eligibility. These requirements are necessary for system security, but can create disproportionate barriers for smaller, distributed, or non-traditional providers when they are not designed in a technology-neutral and performance-based manner. Operational uncertainty surrounding newer technologies may also slow their participation in some service categories. This is reflected, for example, in Great Britain’s recent mid-term stability market, where contracts were awarded mainly to synchronous condensers, pumped-storage hydro assets with zero-MW capability, and synchronous generators with clutch capability, while no contracts were awarded to BESS \cite{neso_as}.
 
\subsubsection{Coordination across energy, balancing, ancillary services, and network operation}
The comparison also shows that no reviewed system fully integrates all ancillary-service requirements within a single market-clearing framework. Texas, California, and Australia coordinate energy and frequency-related services relatively closely through centralised dispatch or co-optimisation, but several non-frequency requirements remain outside these processes. European systems benefit from cross-border energy and balancing integration while continuing to rely on national or zonal arrangements for internal congestion and locational system-security needs. Great Britain similarly coordinates several separate service-specific procurement mechanisms with the Balancing Mechanism. The challenge is therefore not simply whether markets should be separate or co-optimised, but how interactions among energy, balancing, network constraints, and system-security services should be represented and coordinated without losing the specific characteristics of each service.

\subsection{Emerging Trends Across Ancillary-Service Markets}

Against these challenges, four directions of development emerge: faster and more granular frequency services, more explicit stability-related procurement, broader participation by emerging resources, and increasing use of locational and targeted procurement.

\subsubsection{Faster and more granular frequency-control products}
Declining synchronous inertia increases the speed at which frequency can change following disturbances, creating greater demand for fast frequency response and reserve products with shorter activation times. Dynamic frequency-response products in Great Britain, UFFR in Italy, FFR in Denmark, fast-response services in Texas, and very fast FCAS in Australia illustrate this development \cite{neso_as, texas_as, au_as, energinetasdenmarkconditions}. Although their design and procurement differ, these products reflect a common move towards greater differentiation by response speed, sustainment duration, and operational purpose.

\subsubsection{More explicit procurement of stability-related capabilities}
Capabilities such as inertia, SCL, DVS, and system-strength support were historically supplied largely as inherent characteristics of synchronous generation. Their availability can no longer be assumed as synchronous generation is displaced by IBRs. As a result, their operational value is becoming more visible and, in some systems, more explicitly procured. Great Britain has developed dedicated stability procurement frameworks, Germany is moving towards explicit inertia procurement, and Australia uses network-support and transitional arrangements to address system-strength and low-synchronous-generation operation requirements \cite{au_as, neso_stability, germany_inertia}. These developments indicate a gradual transition from implicit provision towards explicit identification and procurement of stability capabilities.

\subsubsection{Broader and more performance-based participation}
Eligibility is increasingly determined by whether a resource can satisfy service-specific performance requirements rather than by technology type alone. This has expanded opportunities for energy storage, demand-side resources, DER aggregations, and appropriately controlled (grid-forming) IBRs. The trend is most developed in frequency-control and reserve markets, where response speed, accuracy, availability, and sustainment can be measured relatively directly. Access remains more selective for voltage support, stability, restoration, and network-support services because their value depends more strongly on location and physical system interaction. Technology neutrality is therefore developing alongside, rather than replacing, service-specific technical qualification.

\subsubsection{Increasing locational and targeted procurement}
Locational procurement is becoming more important for services whose contribution depends strongly on network topology and local operating conditions. Voltage support, system strength, stability services, congestion management, and restoration capability cannot always be valued adequately through system-wide requirements because a resource that is valuable at one location may provide limited benefit elsewhere. Targeted tenders, zonal requirements, network-support arrangements, and other location-specific mechanisms increasingly allow SOs to specify where capability is required. This trend is likely to become more important as renewable generation, storage, and large flexible loads become more geographically concentrated and network constraints increasingly influence system-security requirements.

\begin{figure*}[t]
    \centering
    \includegraphics[width=0.99\textwidth]{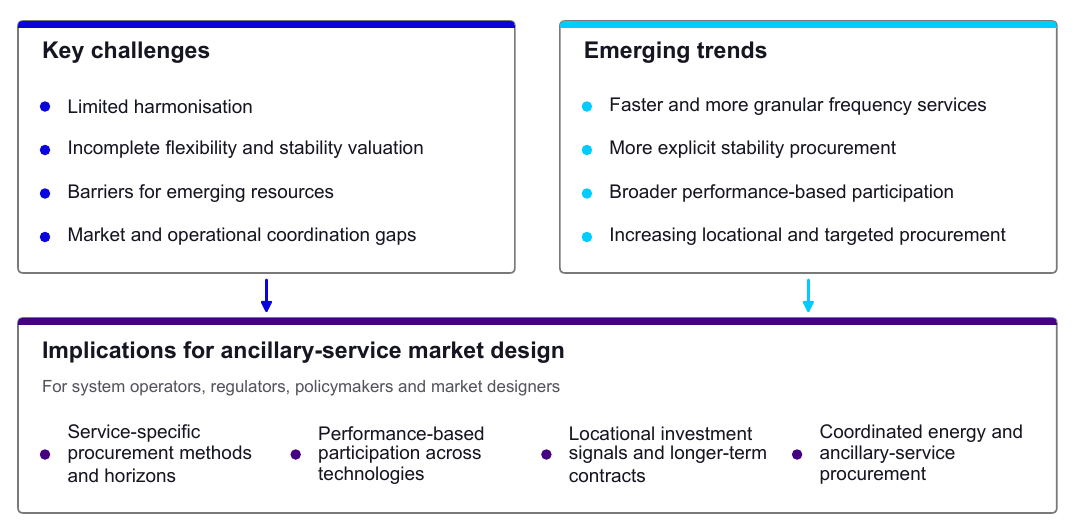}
    \caption{Synthesis of the main challenges and emerging trends identified across the reviewed electricity systems and their implications for ancillary-service market design. Challenges and trends are presented as parallel findings rather than one-to-one relationships.}
    \label{fig:challenges_trends}
\end{figure*}

\medskip
These developments point towards ancillary-service frameworks that are more differentiated by response speed, location, performance, and system-security contribution. Greater differentiation can improve operational precision and reveal previously implicit service needs, but it also creates trade-offs. More granular products can increase complexity and fragment liquidity, while tighter locational requirements can reduce the number of eligible providers. Market evolution therefore requires a balance between precise representation of system needs and sufficiently simple, liquid, and accessible procurement arrangements. Figure~\ref{fig:challenges_trends} synthesises these findings by linking the main challenges and emerging trends identified across the reviewed systems to the broader direction of ancillary-service market design.

\subsection{Market Design and Regulatory Implications}

Building on the synthesis in Figure~\ref{fig:challenges_trends}, system operators, regulators, policymakers, and other actors involved in market design should tailor ancillary-service arrangements to the physical and operational characteristics of each service rather than pursue a single market model. Frequently procured and sufficiently standardised services, particularly frequency control and operating reserves, are generally well suited to recurring auctions, market-clearing mechanisms, and coordination with energy markets. By contrast, services whose value depends strongly on location, specialised technical capability, infrequent system conditions, or longer-term availability may require targeted tenders, bilateral contracts, technical obligations, or longer-term procurement frameworks. The regulatory challenge is therefore to determine which services can be organised effectively as competitive short-term products and which require more tailored arrangements.

Participation rules should similarly reflect the capability required from a resource rather than prescribe a particular technology wherever this is technically feasible. This is particularly important for grid-forming IBRs, DERs, aggregators, and demand-side resources, whose contributions may differ from those of conventional synchronous generators but can nevertheless support secure system operation. A performance-based approach can broaden participation while retaining service-specific requirements. Technology neutrality should therefore not imply identical access conditions across all services, but rather that resources capable of delivering the required performance are not excluded solely because of their technology type.

Investment signals are particularly important for stability and network-support capabilities whose value may arise only at specific locations or under particular system conditions. Such services can be difficult to finance through short-term energy or balancing revenues alone, especially where participation requires new assets or significant upgrades to control systems, protection, communications, or grid-forming functionality. Uncertainty over future service requirements, procurement volumes, contract duration, and remuneration can increase investor risk and the weighted average cost of capital (WACC), potentially discouraging otherwise valuable investments \cite{STEFFEN2020104783,IRENA2023,IEA2024CostCapital}. Longer procurement horizons, transparent assessments of future system needs, clear locational signals, and sufficiently predictable contract structures can reduce part of this uncertainty and improve the investment case for resources such as synchronous condensers and grid-forming IBRs.

Greater coordination between energy and ancillary-service procurement is also desirable, but this does not imply that all services should be incorporated into a single market-clearing process. Co-optimisation can improve efficiency where service requirements, network constraints, opportunity costs, and interactions between competing uses of resource capacity can be represented adequately. This is particularly relevant for energy, frequency response, and reserve services, and may also extend to locational services where their technical and network requirements can be modelled explicitly. However, targeted procurement remains appropriate where service value is highly context-dependent, difficult to standardise, or linked to longer-term capability requirements. Future frameworks are therefore likely to combine co-optimised short-term markets with targeted locational procurement, grid-code obligations, and longer-term contracts where these are better suited to the underlying service.

Procurement horizons, remuneration, participation rules, and market integration should therefore reflect each service's technical requirements, delivery timescales, and locational value. Such differentiated but coordinated arrangements can support broader technology participation while maintaining system security.

\section{Conclusions}
\label{s6}

{\balance This paper reviewed ancillary-service market design across seven electricity systems and five service categories, covering procurement, remuneration, technology participation, and market coordination. Despite common pressures from renewable growth and declining synchronous generation, their frameworks differ because of market architecture, network characteristics, regulation, and operational requirements.

A central finding is that frequency-control and reserve services are the most consistently formalised through explicit products and market-based mechanisms because their requirements are comparatively standardised and measurable. By contrast, voltage and reactive power support, system stability, restoration, and congestion management remain more system-specific, reflecting locational value, network conditions, specialised capabilities, and the operating state of the system. They therefore rely more heavily on targeted tenders, contracts, technical obligations, regulated arrangements, or system-operator actions.

The review also identifies a broader transition in the way ancillary services are defined and provided. Frequency products are becoming faster and more differentiated, stability-related capabilities are being identified and procured more explicitly, participation is becoming increasingly performance-based, and locational procurement is gaining importance. At the same time, energy storage, demand-side resources, DER aggregations, and appropriately controlled IBRs are expanding the range of potential service providers. Their participation, however, remains dependent on service-specific requirements. Market access should therefore become more technology-neutral where technically feasible without weakening the performance requirements needed for secure system operation.

Challenges remain in harmonising non-frequency services, valuing locational and condition-dependent flexibility and stability, overcoming disproportionate participation barriers for emerging resources, and coordinating energy, balancing, ancillary-service procurement, and network operation. Addressing these challenges does not require all services to converge towards a single market structure. Standardised, frequently procured services generally suit short-term competitive markets and co-optimisation, while capabilities dependent on location, specialised equipment, or substantial investment may require targeted procurement, longer-term contracts, or technical obligations.

Future ancillary-service frameworks should therefore combine service-specific arrangements with effective coordination. This can support efficient short-term operation while providing incentives to invest in the flexibility and system-security capabilities needed as renewable generation and other IBRs expand.


{
\onecolumn

\appendix
\section{Appendix}
\label{appendix1}

\newcommand{\cat}[1]{\textsuperscript{\scalebox{0.85}{$#1$}}}

\centering
\scriptsize
\setlength{\tabcolsep}{2.1pt}
\renewcommand{\arraystretch}{1.2}

\begin{longtable}{
C{0.060\textwidth}   
L{0.095\textwidth}   
L{0.135\textwidth}   
C{0.115\textwidth}   
C{0.085\textwidth}   
C{0.080\textwidth}   
C{0.025\textwidth}   
C{0.105\textwidth}   
C{0.080\textwidth}   
C{0.110\textwidth}   
}

\caption{Ancillary-service characteristics by country (SO) and category.}
\label{tab:summary_table}\\

\toprule
\textbf{Country (SO)} &
\textbf{Service} &
\textbf{Operational purpose} &
\makecell{\textbf{Response time}\textsuperscript{2}\\\scriptsize (activation,\\full availability,\\duration)} &
\textbf{Directionality} &
\textbf{Symmetry} &
\textbf{Fault} &
\textbf{Procurement} &
\textbf{Settlement} &
\makecell{\textbf{Payment}\\\textbf{structure}} \\
\midrule
\endfirsthead

\caption[]{Ancillary-service characteristics by country (SO) and category (continued).}\\

\toprule
\textbf{Country (SO)} &
\textbf{Service} &
\textbf{Operational purpose} &
\makecell{\textbf{Response time}\textsuperscript{1}\\\scriptsize (activation,\\full availability,\\duration)} &
\textbf{Directionality} &
\textbf{Symmetry} &
\textbf{Fault} &
\textbf{Procurement} &
\textbf{Settlement} &
\makecell{\textbf{Payment}\\\textbf{structure}} \\
\midrule
\endhead

\endfoot

\bottomrule
\multicolumn{10}{p{0.96\textwidth}}{
\footnotesize
\textit{Category symbols:}
$\cat{\triangle}$ = frequency control and reserves,
$\cat{\bigcirc}$ = voltage and reactive power support,
$\cat{\lozenge}$ = system stability,
$\cat{\square}$ = restoration,
$\cat{\triangledown}$ = congestion management.
\quad
\textit{Directionality:}
$\uparrow$ = upward,
$\downarrow$ = downward,
$\uparrow\downarrow$ = both upward and downward.
\quad
\textit{Symmetry} = symmetric service requirement, reported as Yes/No.
\quad
\textit{Fault} = pre- or post-fault service, reported as Pre/Post.
\par\vspace{0.35em}
\par
\textsuperscript{1}Activation time is relative to the disturbance, while full availability and duration are measured from the activation time ($t+x$).

}\\
\endlastfoot


\multirow{18}{*}[-14em]{%
\centering
\rotatebox[origin=c]{90}{\makecell[c]{Great Britain\\(NESO)\\ \cite{neso_as, neso_stability}}}%
}
& DM\cat{\triangle} & Provides a fast-acting response during volatile periods & 0.5 s / 1 s / 30 min & $\uparrow\downarrow$ & No & Pre & Market-based & Pay-as-clear & Available capability \\
& DR\cat{\triangle}   & Maintains $f$ close to 50 Hz & 2 s / 10 s / 60 min & $\uparrow\downarrow$ & No & Pre & Market-based & Pay-as-clear & Available capability \\
& DC\cat{\triangle}   & Injects or absorbs active power & 0.5 s / 1 s / 15 min & $\uparrow\downarrow$ & No & Post & Market-based & Pay-as-clear & Available capability \\
& SFFR\cat{\triangle} & Injects active power & When $f$<49.7 Hz / 30 s / 30 min & $\uparrow$ & No & Post & Market-based & Pay-as-clear & Available capability \\
& QR\cat{\triangle}   & Provides active power & Instruction / 1 min / 5 min & $\uparrow\downarrow$ & No & Post & Market-based & Pay-as-clear & Available capability, utilisation \\


& SR\cat{\triangle}   & Restores $f$ to $\pm$ 0.2 Hz & Instruction / 100\% ramp-up / 15 min & $\uparrow\downarrow$ & No & Post & Market-based & Pay-as-clear & Available capability, utilisation \\
& BR\cat{\triangle}   & Corrects energy imbalances & Case-specific & $\uparrow\downarrow$ & No & Post & Market-based & Case-specific & Available capability, utilisation  \\
& MFR\cat{\triangle}  & Provides frequency response & Instruction / 10-30 s / 30 min or sustained & $\uparrow\downarrow$ & Yes & Post & Grid code & Pay-as-bid & Available capability, utilisation \\

\addlinespace[0.25em]

& ORPS\cat{\bigcirc}                  & Injects or absorbs reactive power & Instruction / 2 min / sustained & $\uparrow\downarrow$ & No & -- & Grid code & Fixed tariff & MVAr delivered \\
& ERPS\cat{\bigcirc}                  & Injects or absorbs reactive power & Instruction / 2 min / sustained & $\uparrow\downarrow$ & No  & -- & Market-based & Pay-as-bid & Available capability, synchronised capability, utilisation \\
& VNS\cat{\bigcirc}                   & Absorbs reactive power & Instruction / Asset specific (2 sec) / 30 min & $\downarrow$ & No & -- & Market-based & Pay-as-bid & Available capability \\
& Reactive Power Market\cat{\bigcirc} & Under development & -- / -- / -- & -- & -- & -- & -- & -- & -- \\

\addlinespace[0.25em]

& SNS\cat{\lozenge}              & Addresses stability issues & Product specific & $\uparrow\downarrow$ & No & -- & Market-based & Pay-as-bid &  Available capability, utilisation \\
& Stability Market\cat{\lozenge} & Under development & -- / -- / -- & -- & -- & -- & -- & -- & -- \\

\addlinespace[0.25em]

& Restoration\cat{\square} & Restores the system after a blackout & Event / 2 h / 10-72 h & $\uparrow$ & No & Post & Tender/contract-based & Pay-as-bid & Available capability, work contribution \\
& CMIS\cat{\triangledown}        & Reduces the costs for constraint management & Instant trip (<150ms) & $\downarrow$ & No & Post & Tender/contract-based & Pay-as-bid & Arming payment, tripping fee \\
& MW Dispatch\cat{\triangledown} & Manages regional transmission constraints & Instruction / Immediate / 2 min & $\downarrow$ & No & -- & Tender/contract-based & Pay-as-bid & Utilisation \\
& TCMS\cat{\triangledown}        & Manages constraints due to congestion & Location-specific & $\uparrow\downarrow$ & No & -- & Hybrid & Pay-as-bid & Available capability, fixed, utilisation \\

\midrule


\multirow{8}{*}[-2em]{%
\centering
\rotatebox[origin=c]{90}{\makecell[c]{Germany\\(50Hertz, TenneT, Amprion, TransnetBW)\\ \cite{germany_as, as_germany_paper, germany_inertia,flex4fact}}}%
}
& FCR\cat{\triangle}  & Provides primary control reserve & Immediate / 30 s / 15 min & $\uparrow\downarrow$ & Yes & Post & Market-based & Pay-as-clear & Available capability \\
& aFRR (Capacity)\cat{\triangle} & Provides secondary control reserve & 30 s / 5 min / 15 min & $\uparrow\downarrow$ & No & Post & Market-based & Pay-as-bid & Available capability \\
& aFRR (Energy)\cat{\triangle} & Provides secondary control reserve & 30 s / 5 min / 15 min & $\uparrow\downarrow$ & No & Post & Market-based & Pay-as-clear & Utilisation \\
& mFRR (Capacity)\cat{\triangle} & Provides tertiary control reserve & 5 min / 12.5 min / 15 min & $\uparrow\downarrow$ & No & Post & Market-based & Pay-as-bid & Available capability \\
& mFRR (Energy)\cat{\triangle} & Provides tertiary control reserve & 5 min / 12.5 min / 15 min & $\uparrow\downarrow$ & No & Post & Market-based & Pay-as-clear & Utilisation \\

\addlinespace[0.25em]
& Voltage and Reactive Power Support\cat{\bigcirc} & Maintains voltage and reactive power within operational limits & Contract-specific & $\uparrow\downarrow$ & No & -- & Tender/contract-based & Pay-as-bid & Available capability, utilisation \\

\addlinespace[0.25em]

& Inertia\cat{\lozenge} & Limits the RoCoF & Immediate & $\uparrow\downarrow$ & No & -- & Market-based & Fixed-price & Available capability \\

\addlinespace[0.25em]

& Restoration\cat{\square}     & Restores the system after a blackout & Contract-specific & $\uparrow$ & No & Post & Tender/contract-based & Pay-as-bid & Available capability \\

\pagebreak
\multirow{1}{*}[1.5em]{%
\centering
\rotatebox[origin=c]{90}{\makecell[c]{Germany\\(continued)}}%
}

& Redispatch\cat{\triangledown}     & Adjusts active power to relieve physical bottlenecks & Instruction / Immediate (contract terms) / - & $\uparrow\downarrow$ & No & - & Tender/contract-based & Pay-as-clear & Utilisation \\

\addlinespace[0.6em]

\midrule


\multirow{13}{*}[-9em]{%
\centering
\rotatebox[origin=c]{90}{\makecell[c]{Italy\\(Terna) \\ \cite{terna_as, terna_as_7page, italy_grid_code}}}%
}

& FCR\cat{\triangle}  & Provides primary control reserve & Immediate / 30 s / 15 min & $\uparrow\downarrow$ & Yes & Post & Market-based & Pay-as-clear & Available capability \\

& aFRR (Capacity)\cat{\triangle} & Provides secondary control reserve & 30 s / 5 min / 15 min & $\uparrow\downarrow$ & No & Post & Market-based & Pay-as-bid & Available capability \\
& aFRR (Energy)\cat{\triangle} & Provides secondary control reserve & 30 s / 5 min / 15 min & $\uparrow\downarrow$ & No & Post & Market-based & Pay-as-clear & Utilisation \\
& mFRR (Capacity)\cat{\triangle} & Provides tertiary control reserve & 5 min / 12.5 min / 15 min & $\uparrow\downarrow$ & No & Post & Market-based & Pay-as-bid & Available capability \\
& mFRR (Energy)\cat{\triangle} & Provides tertiary control reserve & 5 min / 12.5 min / 15 min & $\uparrow\downarrow$ & No & Post & Market-based & Pay-as-clear & Utilisation \\
& UFFR\cat{\triangle} & Provides very fast frequency control service & Immediate / 1 s / 30 s & $\uparrow\downarrow$ & No & Post & Market-based & Pay-as-clear & Available capability \\

\addlinespace[0.25em]

& Voltage and Reactive Power Support\cat{\bigcirc} & Maintains voltage and reactive power within operational limits & Case-specific & $\uparrow\downarrow$ & -- & -- & Grid code & Contract-specific & Upon-request remuneration \\

\addlinespace[0.25em]

& Inertia\cat{\lozenge} & Limits the RoCoF & Case-specific & $\uparrow\downarrow$ & -- & -- & Grid code & Not provided & -- \\
& SCL\cat{\lozenge} & Contributes to fault support & Case-specific & $\uparrow$ & -- & -- & Grid code & Not provided & -- \\
& Dynamic Oscillation\cat{\lozenge} & Counterattacks oscillations & Case-specific & $\uparrow\downarrow$ & -- & -- & Grid code & Not provided & -- \\

\addlinespace[0.25em]

& Restoration\cat{\square}     & Restores the system after a blackout & Event / 15 min / 24 h & $\uparrow$ & No & Post & Grid code & Pay-as-bid or regulated pricing & Upon activation \\

\addlinespace[0.25em]

& Redispatch (Mixed Injection/ Withdrawal)\cat{\triangledown}     & Adjusts active power to relieve physical bottlenecks & Instruction / Immediate (terms) / - & $\uparrow\downarrow$ & No & - & Grid code & Not provided & - \\

& Extraordinary Instantaneous/ Slow Upward/ Downward Modulation\cat{\triangledown}     & Adjusts the level of active power &  - / - / - & $\uparrow\downarrow$ & No & - & Grid code & - & Regulated settlements \\

\midrule


\multirow{13}{*}[-8em]{%
\centering
\rotatebox[origin=c]{90}{\makecell[c]{Denmark\\(Energinet)\\ \cite{denmark_as, energinetasdenmarkconditions, nordicffr,nordicfcr,  nordicafrr,   nordicmfrr}}}%
}
& FCR\textsuperscript{DK1}\cat{\triangle}  & Provides primary control reserve & Immediate / 30 s / 15 min & $\uparrow\downarrow$ & Yes & Post & Market-based & Pay-as-clear & Available capability \\
& aFRR\textsuperscript{DK1} (Capacity)\cat{\triangle} & Provides secondary control reserve & 30 s / 5 min / continuous & $\uparrow\downarrow$ & No & Post & Market-based & Pay-as-clear & Available capability \\
& aFRR\textsuperscript{DK1} (Energy)\cat{\triangle} & Provides secondary control reserve & 30 s / 5 min / continuous & $\uparrow\downarrow$ & No & Post & Market-based & Pay-as-clear & Utilisation \\
& mFRR\textsuperscript{DK1} (Capacity)\cat{\triangle} & Provides tertiary control reserve & 5 min / 12.5 min / continuous & $\uparrow\downarrow$ & No & Post & Market-based & Pay-as-clear & Available capability \\
& mFRR\textsuperscript{DK1} (Energy)\cat{\triangle} & Provides tertiary control reserve & 5 min / 12.5 min / continuous & $\uparrow\downarrow$ & No & Post & Market-based & Pay-as-clear & Utilisation \\

& FCR-N\textsuperscript{DK2}\cat{\triangle}  & Provides primary control reserve (normal operation) & When $f$<$\pm$100 mHz / 2.5 min / continuous & $\uparrow\downarrow$ & Yes & Pre & Market-based & Pay-as-clear/mFRR energy activation price & Available capability, utilisation \\
& FCR-D\textsuperscript{DK2}\cat{\triangle}  & Provides primary control reserve (under disturbance) & When $f$<$\pm$500 mHz / 7.5 s / continuous & $\uparrow\downarrow$ & No & Post & Market-based & Pay-as-clear & Available capability \\
& aFRR\textsuperscript{DK2} (Capacity)\cat{\triangle} & Provides secondary control reserve & 30 s / 5 min / continuous & $\uparrow\downarrow$ & No & Post & Market-based & Pay-as-clear & Available capability \\
& aFRR\textsuperscript{DK2} (Energy)\cat{\triangle} & Provides secondary control reserve & 30 s / 5 min / continuous & $\uparrow\downarrow$ & No & Post & Market-based & Pay-as-clear & Utilisation \\
& mFRR\textsuperscript{DK2} (Capacity)\cat{\triangle} & Provides tertiary control reserve & 5 min / 12.5 min / continuous & $\uparrow\downarrow$ & No & Post & Market-based & Pay-as-clear & Available capability \\

& mFRR\textsuperscript{DK2} (Energy)\cat{\triangle} & Provides tertiary control reserve & 5 min / 12.5 min / continuous & $\uparrow\downarrow$ & No & Post & Market-based & Pay-as-clear & Utilisation \\
& FFR\textsuperscript{DK2}\cat{\triangle} & Provides fast frequency response & Immediate / 0.7-1.3 s / 5-30 s & $\uparrow$ & No & Post & Market-based & Pay-as-clear & Available capability \\

\addlinespace[0.25em]

& Voltage and Reactive Power Support\cat{\bigcirc} & Stabilises and reconstructs the voltage & Case-specific & $\uparrow\downarrow$ & -- & -- & Grid code & Not provided & -- \\

\pagebreak
\multirow{2}{*}[0em]{%
\centering
\rotatebox[origin=c]{90}{\makecell[c]{Denmark\\(continued)}}%
}

& Restoration\cat{\square}     & Restores the system after a blackout & Contract-specific & $\uparrow$ & No & Post & Tender/contract-based & Pay-as-bid & Available capability \\

\addlinespace[0.25em]

& Upward/ Downward Regulation\cat{\triangledown}     & Avoids overloading the transmission grid
by handling temporary bottlenecks  & - / - / - & $\uparrow\downarrow$ & No & - & Market-based & ``Special regulation'' price & Utilisation \\

\midrule


\multirow{11}{*}[-8em]{%
\centering
\rotatebox[origin=c]{90}{\makecell[c]{Texas\\(ERCOT)\\ \cite{texas_as,ercotas_study, texas_nodal_protocol, ercot_grid_insights}}}%
}
& Reg-Up\cat{\triangle}  & Provides upward reserve capacity & Instruction / 4 s / 30 min & $\uparrow$ & No & Pre & Market-based & Pay-as-clear & Available capability\\
& Reg-Down\cat{\triangle}  & Provides downward reserve capacity & Instruction / 4 s / 30 min & $\downarrow$ & No & Pre & Market-based & Pay-as-clear & Available capability \\
& FRRS-Up\cat{\triangle} & Provides fast upward regulation & Instruction / 1 s / >8 min & $\uparrow$ & No & Pre & Market-based & Pay-as-clear & Available capability \\
& FRRS-Down\cat{\triangle} & Provides fast downward regulation & Instruction / 1 s / >8 min & $\downarrow$ & No & Pre & Market-based & Pay-as-clear & Available capability \\
& PFR\cat{\triangle} & Provides upward dispatchable capacity & When $f$<59.983Hz / 60 s / 30 min & $\uparrow$ & No & Pre & Market-based & Pay-as-clear & Available capability \\
& UFR\cat{\triangle} & Provides upward frequency response via high-set under-frequency relays & When $f$<59.7Hz / 0.5 s / 30 min & $\uparrow$ & No & Post & Market-based & Pay-as-clear & Available capability \\

& FFR\cat{\triangle} & Provides fast upward frequency response & When $f$<59.85Hz / 0.25 s / 15 min & $\uparrow$ & No & Post & Market-based & Pay-as-clear & Available capability \\
& ECRS\cat{\triangle} & Addresses forecast errors, ramping shortages, and reserve replacement & When $f$ < 59.91 Hz or 10-min ramp shortage / 10 min / 1 h & $\uparrow$ & No & Post & Market-based & Pay-as-clear & Available capability  \\
& Non-SP\cat{\triangle} & Addresses low real-time reserves, forecast errors, ramping shortages, and reserve replacement & Physical responsive capability < 3,200 MW or 30-min ramp shortage / 30 min / 4 h  & $\uparrow$ & No & Post & Market-based & Pay-as-clear & Available capability \\

\addlinespace[0.25em]

& Voltage and Reactive Power Support\cat{\bigcirc} & Maintains voltage and reactive power within operational limits  & Case-specific & $\uparrow\downarrow$ & -- & -- & Grid code & Obligation-
based or out-of-market compensation & -- \\

\addlinespace[0.25em]

& Restoration\cat{\square}     & Restores the system after a blackout & Contract-specific & $\uparrow$ & No & Post & Tender/contract-based & Pay-as-bid & Available capability \\
\midrule


\multirow{7}{*}[-7em]{%
\centering
\rotatebox[origin=c]{90}{\makecell[c]{California\\(CAISO)\\ \cite{caiso_as, caiso_bm, caiso_fifth}}}%
}
& Reg-Up\cat{\triangle}  & Corrects supply deficits and maintains frequency near 60 Hz & Instruction / 10 min / continuous & $\uparrow$ & No & Pre & Market-based & Pay-as-clear & Available capability, mileage payment\\
& Reg-Down\cat{\triangle}  & Corrects supply surpluses and maintains frequency near 60 Hz & Instruction / 10 min / continuous & $\downarrow$ & No & Pre & Market-based & Pay-as-clear & Available capability, mileage payment \\
& SP\cat{\triangle} & Provides synchronized reserve capacity after contingencies & Instruction / 10 min / 30 min & $\uparrow$ & No & Post & Market-based & Pay-as-clear & Available capability \\

& Non-SP\cat{\triangle} & Provides non-synchronized reserve capacity after contingencies & Instruction / 10 min / 30 min& $\uparrow$ & No & Post & Market-based & Pay-as-clear & Available capability \\
& FRP\cat{\triangle} & Provides ramping flexibility to manage net-load uncertainty & Instruction / 5 min / 5 or 15 min & $\uparrow\downarrow$ & No & Pre & Market-based & Marginal opportunity cost (Pay-as-clear) & Available capability \\

\addlinespace[0.25em]

& Voltage and Reactive Power Support\cat{\bigcirc} & Maintains voltage and reactive power within operational limits & Case-specific & $\uparrow\downarrow$ & -- & -- & Grid code & Contractual agreements & -- \\

\addlinespace[0.25em]

& Restoration\cat{\square}     & Restores the system after a blackout & Contract-specific & $\uparrow$ & No & Post & Tender/contract-based & Pay-as-bid & Available capability \\
\midrule


\multirow{4}{*}[-3em]{%
\centering
\rotatebox[origin=c]{90}{\makecell[c]{Australia\\(AEMO) \\ \cite{au_as}}}%
}
& RREG\cat{\triangle}  & Corrects supply deficits and maintains frequency near 50 Hz & Instruction / -- / continuous adjustment & $\uparrow$ & No & Pre & Market-based & Pay-as-clear & Available capability \\
& LREG\cat{\triangle}  & Corrects supply surpluses and maintains frequency near 50 Hz & Instruction / -- / continuous adjustment & $\downarrow$ & No & Pre & Market-based & Pay-as-clear & Available capability \\

& Very Fast FCAS Raise (R1)\cat{\triangle} & Arrests frequency decline after contingencies & Event / 1 s / 6 s & $\uparrow$  & No & Post & Market-based & Pay-as-clear & Available capability \\

& Very Fast FCAS Lower (L1)\cat{\triangle} & Arrests frequency rise after contingencies & Event / 1 s / 6 s & $\downarrow$  & No & Post & Market-based & Pay-as-clear & Available capability \\

\pagebreak
\multirow{10}{*}[-8.5em]{%
\centering
\rotatebox[origin=c]{90}{\makecell[c]{Australia\\(continued)}}%
}

& Fast FCAS Raise (R6)\cat{\triangle} & Arrests and stabilises frequency decline after contingencies & R1 / 6 s / 60 s & $\uparrow$ & No & Post & Market-based & Pay-as-clear & Available capability \\
& Fast FCAS Lower (L6)\cat{\triangle} & Arrests and stabilises frequency rise after contingencies & L1 / 6 s / 60 s & $\downarrow$ & No & Post & Market-based & Pay-as-clear & Available capability \\

& Slow FCAS Raise (R60)\cat{\triangle} & Stabilises frequency decline after contingencies & R6 / 60 s / 5 min & $\uparrow$ & No & Post & Market-based & Pay-as-clear & Available capability \\
& Slow FCAS Lower (L60)\cat{\triangle} & Stabilises frequency rise after contingencies & L6 / 60 s / 5 min & $\downarrow$ & No & Post & Market-based & Pay-as-clear & Available capability \\

& Delayed FCAS Raise (R5)\cat{\triangle} & Restores frequency after contingencies & R60 / 5 min / 10 min & $\uparrow$ & No & Post & Market-based & Pay-as-clear & Available capability \\
& Delayed FCAS Lower (L5)\cat{\triangle} & Restores frequency after contingencies & L60 / 5 min / 10 min & $\downarrow$ & No & Post & Market-based & Pay-as-clear & Available capability \\

& RSAS\cat{\bigcirc\lozenge\square\triangledown} & Supports system security, voltage control, network loading, and stability requirements & Case-specific & $\uparrow\downarrow$ & No & -- & Tender/contract-based & Contract-specific & Available capability, enablement, usage, testing \\

\addlinespace[0.25em]

& MBAS\cat{\bigcirc\lozenge\square\triangledown} & Enhances network capability by addressing thermal, voltage, and stability limits & Case-specific & $\uparrow\downarrow$ & No & -- & Tender/contract-based & Contract-specific & Available capability, enablement, usage, testing \\

\addlinespace[0.25em]

& Transitional Services\cat{\lozenge} & Supports emerging system-security needs during the energy transition & Case-specific & $\uparrow\downarrow$ & No & -- & Tender/contract-based & Contract-specific & -- \\

\addlinespace[0.25em]

& SRAS\cat{\square}     & Restores the system after major outages & Contract-specific & $\uparrow$ & No & Post & Tender/contract-based or direct request for offers & Pay-as-bid & Available capability, utilisation, testing \\

\end{longtable}

}


\begin{multicols}{2}
\bibliographystyle{extra/elsarticle-model/model1-num-names}
\bibliography{references}

\end{multicols}

\end{document}